\documentclass[11pt]{article}
\usepackage{graphicx,epsfig} 
\usepackage{amsmath,amssymb,amsfonts,dsfont,amsthm,mathtools,mathrsfs}
\usepackage{bm} 
\usepackage[title]{appendix}

\usepackage{color}
\usepackage[usenames,dvipsnames]{xcolor}
\usepackage[colorlinks=true]{hyperref}
\usepackage{float}
\hypersetup{
    colorlinks = true,
    citecolor = {PineGreen},
}
\usepackage{siunitx}

\def\a{\alpha}
\def\b{\beta}
\def\g{\gamma}
\def\d{\delta}
\def\m{\mu}

\def\n{\nu}

\def\r{\rho}
\def\s{\sigma}

\def\vp{\varphi}

\def\be{\begin{equation}}
\def\ee{\end{equation}}
\def\P{\mathcal{P}}
\def\G{\mathcal{G}}
\def\L{\mathcal{L}}
\def\C{\mathcal{C}}
\def\I{\mathcal{I}}

\def\M{\mathcal{M}}

\def\meff{m_{\text{eff}}}
\def\mp{M_{\text{Pl}}}
\def\D{\mathcal{D}}

\usepackage[vcentermath]{youngtab}
\usepackage{slashed}
\usepackage{textcomp}
\usepackage{ytableau}
\usepackage[bb=boondox]{mathalfa}
\usepackage{tensor}
\usepackage{comment}
\usepackage{setspace}
\usepackage{svrsymbols}
\usepackage{cite}
\usepackage[font=small,skip=0pt]{caption}

\begin{document}

\setlength\arraycolsep{2pt}

\renewcommand{\theequation}{\arabic{section}.\arabic{equation}}
\setcounter{page}{1}

\begin{titlepage}

\begin{center}

\vskip 1.5 cm

{\huge\bf EFT Approach to Black Hole Scalarization and its Compatibility with Cosmic Evolution}

\vskip 2.0cm

{\Large 
Cristi\'an Erices\footnote{cristian.erices@ucentral.cl}$\,^{\spin}$$^{\interaction}$, Sim\'on Riquelme\footnote{sriquelm@ing.uchile.cl}$\,^{\atom}$, and Nicol\'as Zalaquett\footnote{nzalaquett@ludique.cl}$\,^{\bigassumption}$
}

\setcounter{footnote}{0} 
\vskip 0.5cm

{\it $^{\spin}$Universidad Central de Chile, Vicerrector\'ia Acad\'emica, Toesca 1783, Santiago, Chile\\$^{\interaction}$Universidad Cat\'olica del Maule, Av. San Miguel 3605, Talca, Chile\\
$^{\atom}$Grupo de Cosmolog\'ia y Astrof\'isica Te\'orica, Departamento de F\'{i}sica, FCFM, \mbox{Universidad de Chile}, Blanco Encalada 2008, Santiago, Chile\\
$^{\bigassumption}$Physics Ludique Research, Las Urbinas 87, Santiago, Chile}

\vskip 2.5cm

\end{center}

\begin{abstract}
We address the issue of black hole scalarization and its compatibility with cosmic inflation and big bang cosmology from an effective field theory (EFT) point of view. In practice, using a well-defined and healthy toy model which (in part) has been broadly considered in the literature, we consider how higher-order theories of gravity, up to cubic operators in Riemann curvature, fit within this context. Interestingly enough, we find that already at this minimal level, there is a non-trivial interplay between the Wilson coefficients which are otherwise completely independent, constraining the parameter space where scalarization may actually occur. Conclusively, we claim that the EFT does exhibit black hole scalarization, remaining compatible with the inflationary paradigm, and admitting General Relativity as a cosmological attractor.
\end{abstract}

\end{titlepage}
\newpage

\tableofcontents

\section{Introduction}

There is an overwhelming amount of evidence that establishes General Relativity (GR) as the most successful theory describing the gravitational interaction in the so-called ``weak-field'' regime. However, the recent gravitational-wave observations \cite{Abbott1,Abbott2,Abbott3} announce the emergence of a new era in astronomy, which will provide the possibility of testing gravity in the ``strong-field'' regime. This mostly unexplored scenario has motivated new physical grounds where modifications of GR may take place. In this context, there exist theories that exhibit a phenomenon dubbed ``spontaneous scalarization''. This process is indeed a distinctive manifestation of gravitational interactions in the strong-field regime. The earliest of such theories is the Damour and Esposito-Far\`ese (DEF) model \cite{def1,def2}, where scalarization occurs in neutron stars. Neutron stars are able to acquire a non-trivial structure since the theory exhibits non-perturbative deviations from GR only in the strong-field regime. However, spontaneous scalarization was later discovered to happen in other compact objects, such as black holes, in the context of higher-order curvature theories \cite{donevaprl,donevaprl2}.

In the strong-field regime of gravity, the effects of higher-order curvature operators become more significant, leading (generically) to the appearance of ghost degrees of freedom \cite{stelle}. As it is well-known, the Gauss-Bonnet invariant is a purely topological term in four dimensions, but it may become dynamical and ``Ostrogradsky ghosts-free'' when a coupling to a scalar field degree of freedom is considered. This feature is the essence of Einstein Scalar-Gauss-Bonnet (ESGB) gravity, which has become a focus of great interest during the last few years. Several hairy black hole solutions and compact objects have been studied in the four-dimensional ESGB theories \cite{stewart,kanti,torii2,yunes,kleihaus,doneva33}. The phenomenon of spontaneous scalarization requires that the theory admits the GR solution with a trivial scalar field \textit{and} the scalarized hairy black hole solutions (or even interior ones like, e.g., neutron and/or boson stars) \cite{bakopoulos1,bakopoulos2}. Indeed, when the mass of the Schwarzschild black hole is below a certain value, the former becomes unstable in regions of strong curvature \cite{donevaprl}, while the latter emerges as a physically favorable configuration.

Since then, a plethora of coupling functions \cite{Doneva2,herdeiro1,herdeiro2}, as well as several attempts to generalize the spontaneous scalarization framework, have been considered to get a better understanding of the scalarization process. In \cite{herdeiro3}, scalarization is studied within a theory where the scalar field is coupled to the Ricci scalar, while \cite{herdeiro4} proceeds in analogy but under the presence of the Chern-Simons invariant. Scalarized black hole solutions and compact objects in asymptotically flat spacetime of ESGB theories were also investigated in \cite{Doneva3,astefanesei,canate}. Moreover, the spontaneous scalarization of black holes extended to AdS/dS asymptotics were studied in \cite{bakopoulos3,herdeiro5,bakopoulos4,lin}. Interestingly enough, asymptotically AdS black holes allow for connecting scalarization with holographic phase transitions in the dual theory \cite{pap3,brihaye}. Furthermore the authors in \cite{pap4} discuss the spontaneous scalarization in $f(R)$ gravity in the presence of a minimally coupled scalar ﬁeld with a self-interacting potential; while recently, within ESGB theory, spin-induced black hole spontaneous scalarization, which is the outcome of linear tachyonic instability triggered by rapid rotation, has been explored in \cite{kleihaus2,sotiriou4,Doneva4,herdeiro6,kleihaus3,wang}.

Another interesting class of modified gravity theories based on higher-order curvature invariants was constructed in \cite{bueno1} by using cubic contractions of the Riemann tensor. This theory, dubbed ``Einsteinian Cubic Gravity'' (ECG), possesses basic healthiness properties such as: (i) having a spectrum identical to that of Einstein gravity, i.e., the metric perturbation \textit{on a maximally symmetric background} propagates only a transverse massless graviton; (ii) it is neither topological nor trivial in four dimensions, and (iii) it is defined such that it is independent of the number of dimensions. It is well-known that, in general, such terms contribute with fourth-order derivatives of the metric in the field equations. However, as it was shown in \cite{bueno2,mann1,mann2}, the original form of the theory is sufficient to admit spherically symmetric black hole solutions with a second-order differential equation for the metric function. Moreover, an extra cubic contribution, which is trivial for a spherically symmetric black hole ansatz at the level of the field equations, allows one to additionally accommodate a Friedmann-Lema\^itre-Robertson-Walker (FLRW) solution with second-order field equations for the scale factor, leading to a ``purely geometric'' inflationary period \cite{edelstein1}. Black hole solutions were also obtained in \cite{cisternaqtg}, alongside the examination of a well-defined cosmology for modified cubic gravity and higher-order terms such as the quartic, quintic and even an infinite tower of higher-curvature corrections to the Einstein-Hilbert action \cite{edelstein2,Erices:2019mkd}. All these features make this theory physically interesting, and hence, worth of further investigation.

So far curvature-induced scalarization has been explored in scalar-tensor theories where the scalar degree of freedom is excited within the extreme curvature regime through invariants made up of the Riemann tensor. However, up to now as far as the authors have noticed, the theories have not followed a ``strict'' Effective Field Theory (EFT) approach, which has led to hasty conclusions about scalarization of cosmological spacetimes. In this context, the aim of this paper is three-fold. First, we want to go a step further by addressing a scalar-tensor EFT that exhibits curvature-induced scalarization, triggered by a set of suitable invariants made up of Riemann tensor, up to cubic order. Second, it is of interest then to investigate within this framework, how the new operators modify a previously claimed catastrophic instability triggered by quantum fluctuations during the inflationary stage in ESGB theory \cite{Anson:2019uto}. Third, we explore the Big Bang Cosmology (BBC) of the model, and check that GR is indeed a late-time cosmological attractor as experiments seem to demand \cite{Antoniou:2020nax}.

The article is organized as follows: Section \ref{themodel} introduces our model and discusses how it achieves black hole scalarization; Section \ref{pertFLRW} considers the theory of perturbations around a FLRW background and discusses the (in-)stability of such a spacetime geometry; Section \ref{attractor} studies whether (or not) the model has GR as a cosmological attractor; Finally, Section \ref{discandconc} discusses our findings, concludes, and ellaborates on future venues of exploration. Appendices \ref{eftsnut} and \ref{spertds} are very brief overviews of the EFT approach and scalar pertubation theory in dS space, respectively.    

\section{The Model and Black Hole Scalarization}\label{themodel}

We want to consider a theory that exhibits scalarization triggered by higher-order curvature operators (other than the Gauss-Bonnet invariant, which is the one usually considered in the literature) and also satisfies some basic criteria, in order to have a well-posed and healthy gravitational system. Consequently, we construct a model that (i) possesses a spectrum identical to that of Einstein gravity \textit{on a maximally symmetric background}, and (ii) is neither topological nor trivial in four spacetime dimensions.

The purely geometric theory has been constructed in ECG gravity \cite{bueno1} while its generalization to even higher dimensions and operators built up from the curvature, known as ``Generalized Quasi-Topological Gravity'' (GQTG) was studied in \cite{mann2}. We start by recalling the cubic operator $\P$ in ECG theory, which reads
\begin{align}\label{P}
\P = 12\tensor{R}{_\m^\r_\n^\s}\tensor{R}{_\r^\g_\s^\d}\tensor{R}{_\g^\m_\d^\n} + \tensor{R}{_\m_\n^\r^\s}\tensor{R}{_\r_\s^\g^\d}\tensor{R}{_\g_\d^\m^\n} - 12\tensor{R}{_\m_\n_\r_\s}\tensor{R}{^\m^\r}\tensor{R}{^\n^\s} + 8\tensor{R}{^\n_\m}\tensor{R}{^\m_\r}\tensor{R}{^\r_\n},
\end{align}
while the operator $\C$, found in GQTG \cite{mann2}, is given by the combination
\begin{align}
    \C = \tensor{R}{_\m_\n_\r_\s}\tensor{R}{^\m^\n^\r_\d}\tensor{R}{^\s^\d} - \frac{1}{4}\tensor{R}{_\m_\n_\r_\s}\tensor{R}{^\m^\n^\r^\s}R - 2\tensor{R}{_\m_\n_\r_\s}\tensor{R}{^\m^\r}\tensor{R}{^\n^\s} + \frac{1}{2}\tensor{R}{_\m_\n}\tensor{R}{^\m^\n}R.
\end{align}
The latter leads to null contributions in the equations of motion (EOM) when evaluated on a static, spherically symmetric ansatz, a feature that has led some authors to arbitrarily neglect it altogether. However, it has been proven that it is the exact combination $\P - 8\C$, the one that leads to cosmologies with a well-posed initial value problem \cite{edelstein1}. Therefore this is the combination of third-order curvature invariants we shall consider in this paper.

\newpage

In order to explore the phenomenon of scalarization, we must include a scalar field without spoiling the conditions already mentioned. Therefore, we will then consider the dynamical system determined by the action
\begin{align}
     S[\tensor{g}{_\m_\n},\vp] = \int d^4x \sqrt{-g}
\left[\frac{\mp^2}{2}R+ \frac{\alpha}{\mp^2}(\P - 8\,\C) - \frac{1}{2}\tensor{g}{^\mu^\nu}\nabla_\mu\varphi\nabla_\nu\varphi + f\left(\varphi/\mp\right)\,\I+\cdots\right], \label{eq:quadratic}
\end{align}
where $f(\vp/\mp)$ is a dimensionless ``coupling function'' between a canonically-normalized scalar field $\vp$ and a set of curvature invariants given by
\begin{align}
\I = -\beta\,\mp^2 R + \gamma\,\G - \frac{\lambda}{\mp^2}\left(\P-8\C\right),
\end{align}
where $\G$ stands for the well-known Gauss-Bonnet operator $\G \equiv \tensor{R}{_\m_\n_\a_\b}\tensor{R}{^\m^\n^\a^\b} - 4\tensor{R}{_\m_\n}\tensor{R}{^\m^\n} + R^2$. Note that $\G$, being a topological invariant, does not satisfy property (ii) unless it is coupled to the scalar field, so this explains why it is only considered within $\I$. Moreover, $\alpha$, $\beta$, $\gamma$, and $\lambda$ are \textit{dimensionless} coupling constants, which are expected to be $\mathcal{O}(1)$ numbers from an EFT point of view (see Appendix \ref{eftsnut}). Hereafter, for simplicity, we will refer to this theory as ``Scalar-Einsteinian Cubic Gravity'' (SECG).

It should be emphasized that here we are assuming a $\vp \to -\vp$ (discrete) symmetry as well as a $\vp \to \vp + \text{constant}$ (shift) symmetry of the scalar Lagrangian, where the latter is only spoiled by gravitational interactions as given by $f(\vp/\mp)\,\I$ and higher-order operators represented by the ellipsis in \eqref{eq:quadratic}. Note that in this paper we will \textit{not} set the (reduced) Planck scale $\mp = 2.4 \times 10^{18}$ GeV to unity as it is usually done in the literature since we want to keep track of it to easily emphasize its role of being the ultimate EFT cut-off of any gravitational system.

The EOM that stem from extremizing the action $S[\tensor{g}{_\m_\n},\vp] = \int d^4x\sqrt{-g}\,\L$ read
\begin{align}
\tensor{R}{^\alpha^\beta^\rho_\mu}\tensor{P}{_\nu_\rho_\alpha_\beta} + 2\nabla^\alpha\nabla^\beta \tensor{P}{_\alpha_\mu_\nu_\beta} + \frac{1}{2}\nabla_\mu\vp\nabla_\nu\vp + \frac{1}{2}\tensor{g}{_\mu_\nu}\mathcal{L} &= 0,\label{EE}\\
\Box\vp + f_{,\vp}\left(\vp/\mp\right)\,\I &= 0,\label{SE}
\end{align}
where $\nabla_\mu$ is the covariant derivative compatible with the spacetime metric $g_{\mu\nu}$, $\Box\vp \equiv \frac{1}{\sqrt{-g}}\partial_\mu\left(\sqrt{-g}\,\tensor{g}{^\mu^\nu}\partial_\nu\vp\right)$, $f_{,\vp} \equiv \frac{df}{d\vp}$, and  $\tensor{P}{_\alpha_\beta_\mu_\nu}$ is defined as\footnote{We use a normalized antisymmetrization convention $A_{[\mu\nu]}=\tfrac{1}{2}(A_{\mu\nu}-A_{\nu\mu})$.}
\begin{align}
\tensor{P}{_\alpha_\beta_\mu_\nu} &\equiv \frac{\partial\mathcal{L}}{\partial\tensor{R}{^\alpha^\beta^\mu^\nu}}\nonumber\\ 
&= \frac{1}{2}\mp^2, \tensor{g}{_\alpha_{[\mu}}\tensor{g}{_{\nu]}_\beta} + \frac{\a}{\mp^2}\tensor{S}{_\alpha_\beta_\mu_\nu} + f\left(\vp/\mp\right)\left(-\frac{\beta}{2}\mp^2, \tensor{g}{_\alpha_{[\mu}}\tensor{g}{_{\nu]}_\beta} + \gamma\,Q_{\alpha\beta\mu\nu} - \frac{\lambda}{\mp^2}\tensor{S}{_\alpha_\beta_\mu_\nu}\right),
\end{align}
where
\begin{align}
\tensor{Q}{_\alpha_\beta_\mu_\nu} &\equiv 4\tensor{g}{_\b_{[\m}}\tensor{R}{_{\n]}_\a} - 4\tensor{g}{_\a_{[\m}}\tensor{R}{_{\n]}_\b} + 2\tensor{g}{_\alpha_{[\mu}}\tensor{g}{_{\nu]}_\beta}R + 2\tensor{R}{_\alpha_\beta_\mu_\nu},\\
\tensor{S}{_\alpha_\beta_\mu_\nu} &\equiv 4\tensor{R}{_\a_{[\mu}}\tensor{R}{_{\nu]}_\b} + 8\tensor{R}{_\a_\b_\r_{[\m}}\tensor{R}{_{\n]}^\r} + 3\tensor{R}{_\a_\b^\r^\s}\tensor{R}{_\m_\n_\r_\s} + 4\left(2\tensor{R}{^\r_{[\a}}\tensor{R}{_{\b]}_\r_\m_\n} + 9\tensor{R}{_\b_\r_\s_{[\n}}\tensor{R}{_{\m]}^\s^\r_\a} + R\tensor{R}{_\a_\b_\m_\n}\right)\nonumber\\
&+ 2\tensor{g}{_\a_{[\m}}\left(6\tensor{R}{_{\nu]}_\rho}\tensor{R}{_\b^\r} - 2\tensor{R}{_{\n]}_\b}R + \tensor{g}{_{\n]}_\r}\left(\tensor{R}{_\r_\s_\g_\d}\tensor{R}{^\r^\s^\g^\d} - 2\tensor{R}{_\r_\s}\tensor{R}{^\r^\s}\right) + 2\tensor{R}{^\r^\s}\tensor{R}{_{\n]}_\s_\b_\r} - 2\tensor{R}{_{\n]}_\r_\s_\g}\tensor{R}{_\b^\r^\s^\g}\right)\nonumber\\
&+ 4\tensor{g}{_\b_{[\m}}\left(\tensor{R}{_{\n]}_\a}R - 3\tensor{R}{_{\n]}_\r}\tensor{R}{_\a^\r} - \tensor{R}{_{\n]}_\s_\a_\r}\tensor{R}{^\r^\s} + \tensor{R}{_{\n]}_\r_\s_\g}\tensor{R}{_\a^\r^\s^\g}\right).
\end{align}
The EOM for the scalar field fluctuation $\delta\vp \equiv \vp - \vp_0$ is given by
\begin{equation}
\Big[\Box + f_{,\vp\vp}(\vp_0/\mp)\,\I\Big]\delta\vp=0,\label{perteq}
\end{equation}
where $\vp_0$ is the scalar field background, while the d'Alembertian operator and $\mathcal{I}$ are computed in a fixed background. To prove that this theory admits black hole scalarization, we start by noting that the Schwarzschild black hole solution is also a trivial solution of the scalar-tensor cubic theory. This may be achieved by a suitable coupling function satisfying both $f_{,\vp}(0) = 0$ and $f_{,\vp\vp}(0) > 0$. The first condition ensures $\vp_0 = 0$ is a solution of the theory, while the second condition has been understood to be necessary for the emergence of a tachyonic instability in the scalar-Gauss-Bonnet model. It is not difficult, though cumbersome, to strictly prove that the linearized Einstein field equations are the same as in ECG, and therefore property (i) is satisfied, provided the aforementioned conditions are fulfilled. Taking for simplicity $f(x) = \frac{1}{2}x^2$, we may read off from \eqref{perteq} an effective mass squared $\meff^2$ of the form 
\begin{align}
\meff^2 = -f_{,\vp\vp}\left(\vp_0/\mp\right)\,\I = \beta R - \frac{\gamma}{\mp^2}\G + \frac{\lambda}{\mp^4}\left(\P - 8\C\right).\label{meff2}
\end{align}
It so happens that a black hole is a solution with $R = 0$ and $\G > 0$, so we clearly see that $\gamma$ must be positive $\left(\gamma > 0\right)$ in order to have a tachyonic instability $\left(\meff^2 < 0\right)$, as the presence of the cubic term should and will be taken to be immaterial because it is further suppressed by the Planck scale for ``natural'' values of $\lambda$. Therefore, in this article we shall always assume 
\begin{align}\label{condgamma}
    \gamma > 0,
\end{align}
as we are interested in scalarized black hole solutions. Let us now consider perturbations on a fixed Schwarzschild background. The symmetry of such a spacetime allows for a decomposition of the perturbation using separation of variables, meaning 
\begin{equation}\label{pert}
\delta\vp = \frac{u(r)}{r}e^{-i\omega t}\,\tensor{Y}{_l_m}(\theta,\phi),
\end{equation}
where $\tensor{Y}{_l_m}(\theta,\phi)$ are the usual spherical harmonics. After substitution of \eqref{pert} into the EOM \eqref{perteq}, and using tortoise coordinates defined through $dr_* = dr\left(1 - \frac{r_g}{r}\right)^{-1}$, with $r_g \equiv \M/4\pi\mp^2$ standing for the Schwarzschild radius of the black hole of mass $\M$, we obtain a ``Schr\"{o}dinger-like'' equation of the form
\begin{align}
\frac{d^2u}{dr_{*}^{2}} + \omega^{2}u = V_\text{eff}(r)u,
\end{align}
where the effective potential $V_\text{eff}$ is given by
\begin{align}
   V_\text{eff}(r) = \left(1 - \frac{r_g}{r}\right)\left(\frac{l(l + 1)}{r^2} + \frac{r_g}{r^3} - \frac{\gamma}{\mp^2}\frac{12\,r_g^2}{r^6} + \frac{\lambda}{\mp^4}\frac{84\,r_g^3}{r^9}\right).
\end{align}
There exists a sufficient condition for the existence of an unstable mode given by \cite{doi:10.1119/1.17935,donevaprl}
\begin{align}
\int_{-\infty}^{\infty}dr_{*}\,V_\text{eff}(r_{*}) = \int_{r_{g}}^{\infty}dr\,\frac{V_\text{eff}(r)}{\left(1 - \frac{r_{g}}{r}\right)} < 0\ .
\end{align}
For this condition to hold, spherically symmetric perturbations $(l = m = 0)$ in a Schwarzschild background require
\begin{align}
    5\,(r_g\mp)^4 - 24\,(r_g\mp)^2\,\gamma + 105\,\lambda < 0.\label{chareq}
\end{align}
This implies that
\begin{align}
\left(\frac{\M}{\mp}\right)^2&\in\left[ q_{-},q_{+}\right]& \text{if}\quad&0<\lambda\leq\frac{48}{175}\,\gamma^2,\label{range2}\\
\left(\frac{\M}{\mp}\right)^2&\in \left[0,q_{+}\right]&\text{if}\quad&\lambda\leq 0,\label{range}
\end{align}
where we have defined
\begin{align}
q_{\pm} \equiv \frac{16\,\pi^2}{5}\left(12\,\gamma \pm \sqrt{144\,\gamma^{2} - 525\,\lambda}\right).
\end{align}

Interestingly enough, the Schwarzschild background is unstable for a specific range of masses given by the above bounds. Moreover, we see that scalarization may only occur when
\begin{align}
    \lambda\leq \frac{48}{175}\,\gamma^2,\label{constcoup}
\end{align}
which is a non-trivial constraint between the couplings. This is still compatible with $\gamma$ and $\lambda$ both being $\mathcal{O}(1)$ numbers. However, the sign of $\lambda$ is clearly not fixed by this condition. Note that when $\lambda = 0$, equations \eqref{chareq} and \eqref{range} imply 
\begin{equation}
 \M^2 < \frac{384\,\pi^2}{5}\,\gamma\,\mp^2.\label{naivebound}   
\end{equation}
A back-of-the-envelope calculation shows then that within our naive quadratic theory (endowed with $\mp$ as the \textit{only} relevant scale), the maximum mass of Schwarzschild black holes that may be scalarized is of order $10^{-37}$ solar masses. This fact certainly precludes any possibility of such a version of SECG theory to be compared with observations. In the next section, we will modify the theory in order to solve this and some other ``problems'' that we will find along the way. Additionally, it is crucial to stress that if we want GR solutions to be admissible in the model we have been considering, we need to check that $\vp = \vp^{(0)} = 0$ is the asymptotic value that $\vp$ needs to take for unscalarized configurations. Let us now consider how scalarization may occur within a cosmological setting.

\section{Perturbations on a FLRW Background}\label{pertFLRW}

In a FLRW background with metric
\begin{equation}
ds^2 = \tensor{g}{_\m_\n}dx^\m dx^\n = -dt^2 + a(t)^2\tensor{\delta}{_i_j}dx^idx^j,\quad \text{and}\quad H\equiv\frac{\dot{a}}{a},
\end{equation}
it so happens that the EOM for the fluctuation reads
\begin{align}
    \delta\ddot{\vp} + 3H\delta\dot{\vp} - \frac{\nabla^2\delta\vp}{a^2} + \meff^2\,\delta\vp = 0,
\end{align}
which, by Fourier expanding $\delta\vp \sim \int d\omega\,d^3\boldsymbol{k}\,\delta\vp(\omega,\boldsymbol{k})\,e^{-i(\omega t - \boldsymbol{k}\cdot\boldsymbol{x})}$, implies that
\begin{align}
    \omega^2 = \frac{k^2}{a^2} + \meff^2,
\end{align}
when neglecting the slow change of $\omega$ on the time scales shorter than $H^{-1}$. Moreover, for a FLRW spacetime
\begin{align}
    R = 6\left(2H^2 + \dot{H}\right),\quad \G = 24H^2\left(H^2 + \dot{H}\right),\quad \P - 8\C = -48H^4\left(2H^2 + 3\dot{H}\right),
\end{align}
so that the effective mass squared becomes
\begin{align}
    \meff^2 = 6\beta\left(2H^2 + \dot{H}\right) - \frac{24\,\gamma}{\mp^2}H^2\left(H^2 + \dot{H}\right) - \frac{48\,\lambda}{\mp^4} H^4\left(2H^2 + 3\dot{H}\right).\label{effmass}
\end{align}
Let us entertain for a second the case $\beta = \lambda = 0$. The effective mass would then read
\begin{align}
\meff^2 = -\frac{24\,\gamma}{\mp^2}H^2\left(H^2 + \dot{H}\right) = -\frac{24\,\gamma}{\mp^2}H^2\frac{\ddot{a}}{a},
\end{align}
so that, as long as $\gamma > 0$,
\begin{align}
    \meff^2 < 0 \Longleftrightarrow \ddot{a} > 0,
\end{align}
implying the existence of a tachyonic instability during any (quasi)-de Sitter (dS) phase of our universe, as it is emphasized in \cite{Anson:2019uto} ($\gamma = 1$ for them). However, we see \textit{no reason}, on EFT grounds, to arbitrarily set $\beta = \lambda = 0$. Moreover, as previously noted, we expect that all $\beta$, $\gamma$, and $\lambda$ are $\mathcal{O}(1)$ numbers, and yet, it is still the case that the dimensionless ratio $\left(H/\mp\right)^2$ does generate a clear hierarchy among the contributions to the effective mass in \eqref{effmass}. Thus, the appropriate thing to do is to consider the effective mass, which by using the standard definition $\epsilon \equiv -\dot{H}/H^2$ we now write 
\begin{align}
    \meff^2 = 12\left[\beta\left(1 - \frac{\epsilon}{2}\right) - 2\,\gamma\left(1-\epsilon\right)\zeta - 8\,\lambda\left(1-\frac{3}{2}\,\epsilon\right)\zeta^2\right]H^2,\label{effmass2}
\end{align}
as an expansion in the small ratio
\begin{align}
\zeta \equiv \left(\frac{H}{\mp}\right)^2.
\end{align}
Say we keep only the zeroth-order term in such an expansion in $\zeta$ so that
\begin{align}
    \meff^2 = 12\beta\left(1 - \frac{\epsilon}{2}\right)H^2.
\end{align} 
We see that since during inflation $\epsilon \ll 1$, the overall sign of $\meff^2$ is simply the sign of $\beta$ itself. If we now consider the leading non-trivial contribution in the expansion in $\zeta$, the effective mass squared may in principle be negative as long as
\begin{align}
    \beta\left(1 - \frac{\epsilon}{2}\right) < 2\,\gamma\left(1 - \epsilon\right)\zeta.
\end{align}
For inflation we may safely take $H \sim 10^{13}$ GeV, so we may assume that $\zeta \sim 10^{-11}$ (strictly positive), implying that
\begin{align}
    \frac{\beta}{\gamma} \lesssim \mathcal{O}(10^{-11}).
\end{align}
We see that if $\beta > 0$ this condition implies a big hierarchy between $\beta$ and $\gamma$, which is clearly not expected nor justified from an EFT perspective. Instead, if $\beta < 0$, the condition may be naturally satisfied. Again then, the possibility of developing a tachyonic instability during inflation is ultimately related to the sign of $\beta$. In other words, it is still the case that
\begin{align}
    \meff^2 < 0 \Longleftrightarrow \beta < 0.
\end{align}
If we assume (for the sake of simplicity) $\beta > 0 $, $\gamma = 0$, $\lambda > 0$, requiring the cubic operator contribution to render a negative effective mass squared would imply that
\begin{align}
    \frac{\beta}{\lambda} \lesssim \mathcal{O}(10^{-22}),
\end{align}
which of course worsens the prospects for the naive idea that higher-order operators may play some non-trivial role on the phenomenon of scalarization during inflation. At the risk of being a little pedantic, we repeat; the sign of $\beta$ univocally determines the fate of scalarization in this weakly broken shift-symmetric theory.

For a dS solution $\epsilon = 0$ so
\begin{align}
    \meff^2 = \left(12\beta - 24\gamma\,\zeta - 96\lambda\,\zeta^2\right)H^2.\label{effmass3}
\end{align}
Let us come back for a (second) second to the $\beta = \lambda = 0$ case. Doing so allows us to estimate the corresponding would-be $\gamma$ constant in the model discussed by the authors of \cite{Anson:2019uto}. Taking, as they do,\footnote{Here, $\bar{\lambda}$ stands for $\lambda$ in \cite{Anson:2019uto} and has nothing to do with our own $\lambda$ introduced in this paper.} $\bar{\lambda} \sim 10^{19}\,\text{GeV}^{-1}$, this would imply a colossal value for $\gamma = \frac{1}{4}\mp^2\bar{\lambda}^2$, namely
\begin{align}
    \gamma \sim 10^{74}.\label{gammaest}
\end{align}
Sweeping this issue under the rug for the time being, one may go ahead and estimate the ratio of the instability time $t_\text{inst}$ to the age of the universe $t_0 \sim 1/H_0$ when taking $H_0 \simeq 10^{-43}$ GeV,
\begin{align}
    \frac{t_\text{inst}}{t_0} \sim \frac{H_0}{\meff} \sim \frac{1}{2\sqrt{6\,\gamma}}\frac{\mp}{H_0} \sim 10^{23},\label{inst1}
\end{align}
and we learn that the instability is not noticeable during current dark energy domination. Instead, during inflation, the estimation delivers
\begin{align}
    \frac{t_\text{inst}}{t_\text{inf}} \sim \frac{1}{N}\frac{H_\text{inf}}{\meff} \sim \frac{1}{2\sqrt{6\,\gamma}N}\frac{\mp}{H_\text{inf}} \sim 10^{-34},\label{inst2}
\end{align}
where $N \sim 10^2$ is the required number of e-folds to overcome the classical shortcomings of BBC. Equation \eqref{inst2} informs us of a very short instability time, implying a catastrophic instability of the theory during inflation, unless the initial values of the field are finely tuned \cite{Anson:2019uto}. Nevertheless, it should be common wisdom that any perturbative scheme around an \textit{unstable} ground state is doomed to exhibit ``runaway'' behavior. The quantum analysis of such finely tuned initial conditions system was indeed made in \cite{Anson:2019uto}, with foreseeable results. However, before moving forward to the analisys of the more sensible theory where $\beta \neq 0$, we should address the issue of getting such a big order of magnitude for the estimation of $\gamma$ in \eqref{gammaest}.

Say we instead had chosen from the onset a function of the form $f\left(\vp/M\right)$ where $M$ stands for some \textit{other} additional scale that we may fix later on. Now the effective mass reads
\begin{align}
\meff^2 = -f_{,\vp\vp}\left(\vp_0/M\right)\,\I = \beta\frac{\mp^2}{M^2}R - \frac{\gamma}{M^2}\G + \frac{\lambda}{M^2\mp^2}\left(\P - 8\C\right),\label{effmassnew}
\end{align}
or
\begin{align}
    \meff^2 &= 6\beta\frac{\mp^2}{M^2}\left(2H^2 + \dot{H}\right) - \frac{24\,\gamma}{M^2}H^2\left(H^2 + \dot{H}\right) - \frac{48\,\lambda}{M^2\mp^2} H^4\left(2H^2 + 3\dot{H}\right)\nonumber\\
    &= 12\left[\beta\left(1 - \frac{\epsilon}{2}\right)\frac{\mp^2}{M^2} - 2\,\gamma\left(1-\epsilon\right)\frac{H^2}{M^2} - 8\,\lambda\left(1-\frac{3}{2}\,\epsilon\right)\frac{H^4}{M^2\mp^2}\right]H^2.\label{effmass4}
\end{align}
Let us then consider one last time the $\beta = \lambda = 0$ case. For scalarization to be relevant for astrophysical black holes we need to pick $M = L^{-1}$, with $L$ being the characteristic length-scale of the compact object which is usually taken to be $L \sim 10$ km., thus arriving at
\begin{align}\label{mscale}
    M = 1.98 \times 10^{-20}\,\text{GeV} \xRightarrow{\gamma = \frac{1}{4}M^2\bar{\lambda}^2} \gamma \sim 10^{-2},
\end{align}
which is a much more sensible number than the one estimated in \eqref{gammaest}. There is still a pending issue though. Staring at \eqref{effmassnew} it is clear that the coefficient accompanying the Ricci scalar, ${\tilde{\beta} \equiv \beta\,\mp^2/M^2}$, cannot be an order one number unless $\beta$ is tuned to be extremely small because the ratio $\mp^2/M^2 \sim 10^{76}$. This implies we would be dealing with a field of super-Planckian mass, which is clearly off the EFT description and makes no sense. The way out of this conundrum is to acknowledge the fact that the mass scale should not only be modified within the coupling function $f$ but within the curvature invariant $\I$ as well, so we should really take
\begin{align}
\I = -\beta\,M^2 R + \gamma\,\G - \frac{\lambda}{M^2}\left(\P-8\C\right),
\end{align}
which implies that the effective mass is finally given by the expression in \eqref{effmass4} after carrying out the replacement $\mp \to M$, so that
\begin{align}
    \meff^2 &= 12\left[\beta\left(1 - \frac{\epsilon}{2}\right) - 2\,\gamma\left(1-\epsilon\right)\chi - 8\,\lambda\left(1-\frac{3}{2}\,\epsilon\right)\chi^2\right]H^2,\quad \text{with} \quad \chi \equiv \left(\frac{H}{M}\right)^2.\label{effmass5}
\end{align}
It is quite interesting that the introduction of the new energy scale $M$ in the problem, besides the ``kinematical'' ($\mp$) and the ``dynamical'' ($H$) ones, ``naturalizes'' an otherwise finely tuned EFT. Moreover, such an energy scale corresponds to nothing but the actual physical extension of the object to become scalarized. We will have more to say about this later. 

With the emergence of the scale $M$, a Schwarzschild black hole of mass $\mathcal{M}$ may be now scalarized provided
\begin{align}
\left(\frac{\M}{M}\right)^2 &\in \left[q_{-}, q_{+}\right] &\text{if}\quad &0 < \lambda \leq \frac{48}{175}\,\gamma^2,\label{range22}\\
\left(\frac{\M}{M}\right)^2 &\in \left[0, q_{+}\right] &\text{if} \quad &\lambda \leq 0,\label{range11}
\end{align}
with $q_{\pm} \equiv \frac{16\,\pi^2}{5}\left(12\, \gamma\pm\sqrt{144\,\gamma^{2} - 525\,\lambda}\right)$ as before. We see now that when $\lambda = 0$, 
\begin{equation}
\M^2 < \frac{384\,\pi^2}{5}\,\gamma\left(\frac{\mp^4}{M^2}\right),\label{truebound}
\end{equation}
which is the bound found in the ESGB theory, up to normalization \cite{donevaprl}. The difference between the inequalities in \eqref{naivebound} and \eqref{truebound} is crucial, since the former would imply that only black holes with extremely small masses may be scalarized, and these are definitely not observable. The introduction of the mass scale $M$ given by \eqref{mscale} makes it possible for black holes of up to 180 solar masses to become scalarized. Hereafter, we will refer to SECG as the theory with this new scale incorporated.

In reference \cite{donevaprl}, the authors consider a positive coupling constant of the Gauss-Bonnet operator, a condition that we also demand in \eqref{condgamma} in order for scalarization of Schwarzschild black holes to become attainable. However, and unlike ESGB, the instability of Schwarzschid black holes is not only triggered once the curvature at the horizon reaches a minimum value within our model, as it must also \textit{not} exceed an upper bound. Such bounds are determined by the coupling constant $\lambda$ associated with the cubic operators in the action. This means that black holes with extremely large curvature are not prone to become scalarized. This last assertion may be quantified by computing the Kretschmann scalar at the horizon $\mathcal{K}_H$, so that the above relations imply that scalarization will occur only when
\begin{align}
\frac{\mathcal{K}_H}{\mp^4} &\in \left[\frac{3072\,\pi^4}{q_{+}^2}, \frac{3072\,\pi^4}{q_{-}^2}\right] &\text{if} \quad &0 < \lambda \leq \frac{48}{175}\,\gamma^2,\label{range222}\\
\frac{\mathcal{K}_H}{\mp^4} &\in \left[0, \frac{3072\,\pi^4}{q_{-}^2}\right] &\text{if} \quad &\lambda \leq 0.\label{range111}
\end{align}

Finally, despite the fact that the estimation of the instability times  in \eqref{inst1} and \eqref{inst2} remain the same, the conclusion that present-day acceleration is safe from the instability, but inflation may not be, is precipitous. Indeed, as we have been firmly trying to emphasize, there are no compelling reasons for having a vanishing $\beta$ value. Let us consider the leading term in \eqref{effmass4}, meaning
\begin{align}
    \meff^2 = 12\beta H^2,
\end{align}
where we have taken the dS approximation, $\epsilon = 0$.
We then find that
\begin{align}
    \frac{t_\text{inst}}{t_0} \sim \frac{H_0}{\meff} \sim \frac{1}{\sqrt{12\,|\beta|}} \sim \frac{0.29}{\sqrt{|\b|}},\label{inst3}
\end{align}
which is marginally different than one, unless $\b$ is tuned to \textit{not} be an order one number. Moreover, in the case of inflation, with $N \sim 10^2$, the relevant ratio reads 
\begin{align}
    \frac{t_\text{inst}}{t_\text{inf}} \sim \frac{1}{N}\frac{H_\text{inf}}{\meff} \sim \frac{1}{\sqrt{12\,|\beta|}N} \sim \frac{2.9\times 10^{-3}}{\sqrt{|\b|}}.\label{inst4}
\end{align}
The estimation in \eqref{inst3} seems to be the end of the story for us, as present-day cosmology would be largely unstable under scalarization within our EFT. Understanding that there is complete loss of generality in arbitrarily setting $\beta = 0$, and the expected hierarchy in the set of curvature operators, has led to the conclusion that in order to have good stable cosmological evolution we need to take $\beta$ \textbf{strictly positive}. In other words, there is \textit{no} chance of having a tachyonic instability during \textit{any} (quasi)-dS stage of the cosmic evolution of our universe, and because of this, we will only consider the case 
\begin{align}
\beta > 0
\end{align}
in the rest of this paper. Let us now check if our SECG model has the nice and desirable property of having GR as a late-time cosmic attractor.

\section{General Relativity as a Cosmic Attractor}\label{attractor}

The scalar EOM \eqref{SE} in a FLRW background is given by
\begin{align}
    \ddot{\vp} + 3H\dot{\vp} + \meff^2\,\vp = 0,\label{SEM2}
\end{align}
where
\begin{align}
    \meff^2 = \beta R - \frac{\gamma}{M^2}\G + \frac{\lambda}{M^4}\left(\P - 8\C\right),
\end{align}
and depends on the cosmological background. As usual, to study the evolution of the scale factor we concentrate in the ``time-time'' component of the modified Einstein equations
\begin{align}
    \mp^2\tensor{G}{_t_t} = \rho_\text{eff} + \rho_a,
\end{align}
where $\rho_a$ denotes the energy densities of the several BBC components of the cosmic fluid, while $\rho_\text{eff}$ denotes an effective energy density associated with the presence of the cubic operator and the scalar field, and it is given by \begin{align}
    \rho_\text{eff} &\equiv \rho_{\P\C} + \rho_\vp,\\ 
    \rho_{\P\C} &= -\frac{48\,\alpha}{\mp^2}H^6,\\
    \rho_\vp &= \frac{1}{2}\dot{\vp}^2 + 6\left(\beta - 4\,\gamma\,\chi - 24\,\lambda\,\chi^2\right)H\vp\dot{\vp} + 3\left(\beta + 8\,\lambda\,\chi^2\right)H^2\vp^2.\label{energydensity}
\end{align}
We shall demand usual cosmic evolution, meaning
\begin{align}
\rho_a \approx 3\mp^2H^2,\label{friedeq} \end{align}
which directly implies that
\begin{align}
|\rho_{\P\C}| \ll \rho_a,
\end{align}
given the expectation (due to EFT reasoning) that $\alpha \ll \left(\mp/H\right)^4$. In fact, notice that even during inflation, which is the cosmic stage when $H$ attains its highest possible value, ${\left(\mp/H\right)^4 \sim 10^{21} \gg \a}$. In other words, the cubic energy density is indeed negligible when compared with BCC cosmic fluid densities. Moreover, as we do not want $\vp$ to play any role in late-time cosmology, we shall assume, for the time being, that 
\begin{align}
\rho_\varphi \ll \rho_a,
\end{align}
though, we acknowledge, it will be mandatory to check if such an assumption is dynamically consistent. Using the Friedmann equation \eqref{friedeq} and the continuity equation
\begin{align}
\dot{\rho}_a + 3H\rho_a\left(1 + w_a\right) = 0,\quad \text{where} \quad w_a \equiv \frac{\rho_a}{p_a},
\end{align}
we may put the curvature invariants in the useful form
\begin{align}
    R &= 6\left(2H^2 + \dot{H}\right) = \frac{\rho_a}{\mp^2}\left(1 - 3\,w_a\right),\\
    \G &= 24H^2\left(H^2 + \dot{H}\right) = -\frac{4}{3}\left(\frac{\rho_a}{\mp^2}\right)^2\left(1 + 3\,w_a\right),\\
    \P - 8\C &= -48H^4\left(2H^2 + 3\dot{H}\right) = \frac{8}{3}\left(\frac{\rho_a}{\mp^2}\right)^3\left(\frac{5}{3} + 3\,w_a\right).
\end{align}
We are interested in the behavior of the scalar field during different cosmological stages. In Table \ref{table1} below we summarize the signs of the Ricci scalar $R$, the Gauss-Bonnet operator $\G$, and the cubic operators $\P - 8\C$. 
\begin{table}[h!]
\centering
\begin{tabular}{||c c c c||} 
\hline
& Radiation & Matter & Dark Energy \\ [0.5ex] 
\hline\hline
$\P-8\C$ & $> 0$ & $> 0$ & $< 0$ \\ 
$\G$ & $< 0$ & $< 0$ & $> 0$ \\
$R$ & $0$ & $> 0$ & $> 0$ \\ [1ex] 
\hline
\end{tabular}
\caption{Signs of the Ricci scalar, the Gauss-Bonnet invariant, and the cubic invariants during different cosmological eras.}
\label{table1}
\end{table}

\newpage

These signs, together with the signs of the dimensionless coupling constants $\beta$, $\gamma$, and $\lambda$, control the sign of the effective mass. The fact that the aforementioned operators have different mass dimension implies that they will scale differently with time. The cubic operator is then naively expected to be dominant during early times.
Considering the expression for the effective mass in \eqref{effmass5}, we see that during radiation domination (RD), $\epsilon = 2$, and 
\begin{align}
    m_\text{eff,R}^2 = 24\,\chi\left(\gamma + 8\,\lambda\,\chi\right)H^2.\label{meffRD}
\end{align}
Staring at \eqref{meffRD} we observe that during RD the dynamics of the scalar field is determined by the higher-order curvature operators, as it is not sensitive to $\beta$, and we will explicitly see such behavior once we numerically solve the scalar field equation. Moreover during matter domination (MD), $\epsilon = \frac{3}{2}$, and one finds that
\begin{align}
    m_\text{eff,M}^2 = 3\left(\beta + 4\,\gamma\,\chi + 40\,\lambda\,\chi^2\right)H^2.
\end{align}
Throughout MD it so happens that $H \ll M$, so we may confirm that, within that era, the contributions from the Gauss-Bonnet and cubic operators are largely subdominant, as expected from dimensional analysis. Finally during dark energy domination (DED), $\epsilon = 0$, so
\begin{align}
    m_\text{eff,DE}^2 = 12\left(\beta - 2\,\gamma\,\chi - 8\,\lambda\,\chi^2\right)H^2,
\end{align}
and since during this stage $H \lll M$, it is clear that the ``destiny of the spacetime itself'' relies on the sign of $\beta$. Luckily for us, having taken the positive option, our current cosmology is safe.

Actually, it has been pointed out in \cite{Antoniou:2020nax}, that the strict $\gamma = \lambda = 0$ limit is intimately related with a linearized version of the seminal DEF model for scalarization \cite{def2}, which has been proven, for $\beta > 0$, to have GR as a cosmological attractor \cite{def1}. As non-linearities are not important for this analysis, it is expected that once the $\G$ and $\P - 8\C$ operators become negligible, the cosmic evolution will match that of the DEF model, meaning the scalar field will be naturally driven to $\vp_0 = 0$. As the transition from MD to DED does not qualitatively change the dynamics of $\vp$ and the influence of higher-order operators becomes more and more marginal, GR with $\vp_0 = 0$ is expected to be an attractor.

Let us verify all the above assertions through a quantitative study of the scalar field dynamics by expressing the scalar EOM in terms of the redshift $z$ instead of cosmic time $t$, so that the scalar field equation now reads
\begin{equation}\label{seqz}
\vp_{(a)}^{\prime\prime}(z) + f_a(z)\vp_{(a)}^{\prime}(z) + q_a(z)\vp_{(a)}(z) = 0,
\end{equation}
where
\begin{eqnarray}
f_{a}(z) &\equiv& \frac{H^{\prime}(z)}{H(z)} - \frac{2}{z+1},\\
q_{a}(z) &\equiv& \frac{3}{(1+z)^2}\left[\beta\left(1 - 3\,\omega_a\right) + 4\,\gamma\,\chi(z)\left(1 + 3\,\omega_a\right) + 8\,\lambda\,\chi(z)^2\left(5 + 9\,\omega_a\right)\right],
\end{eqnarray}
the primes denote differentiation with respect to $z$, $\vp'(z) \equiv \frac{d\vp(z)}{dz}$, and ${\chi(z) \equiv \left(H(z)/M\right)^2}$. We note that the form of the Hubble friction and ``mass'' terms $f_a(z)$ and $q_a(z)$ depend on the effective energy density that drives the cosmic evolution. Explicitly, we have that during
\begin{align}
    &\text{RD:}&\quad f_\text{r}(z)& = 0,&\quad q_\text{r}(z)& = \frac{24 }{(1+z)^2}\,\chi(z)\left[\gamma + 8\,\lambda\,\chi(z)\right],\quad &\omega_\text{r} &= \frac{1}{3},\label{qr}\\
    &\text{MD:}&\quad f_\text{m}(z)& = -\frac{1}{2\left(1 + z\right)},&\quad q_\text{m}(z)&= \frac{3}{(1+z)^2}\left[\beta + 4\,\gamma\,\chi(z) + 40\,\lambda\,\chi(z)^2\right],\quad &\omega_\text{m} &= 0,\\
    &\text{DED:}&\quad f_\text{m}(z)&= -\frac{2}{1 + z},&\quad q_\text{de}(z)&=\frac{12}{(1+z)^2}\left[\beta - 2\,\gamma\,\chi(z) - 8\,\lambda\,\chi(z)^2\right],\quad &\omega_\text{de} &= -1.
\end{align}
The numerical analysis is set to start at $z_i = 10^{10}$, right before big bang nucleosynthesis (BBN) epoch. A natural initial value for the dimensionless field\footnote{Incidentally, such a field redefinition recovers the quadratic Lagrangian considered in \cite{Antoniou:2020nax}.} $\tilde{\vp} \equiv \vp/\sqrt{2}\mp$ is given by 
\begin{align}
\tilde{\vp}_i = \frac{\vp_i}{\sqrt{2}\mp} \simeq \frac{H}{\sqrt{2}\mp} \ll 1,\label{phiini}
\end{align}
as $\vp$ is basically coupled with the thermal bath. Moreover, it is also natural to expect that $\dot{\tilde{\vp}}_i\simeq H(z_i)\tilde{\vp}_i$, which implies that 
\begin{align}
|{\tilde{\vp}_i}'| \simeq \left|\frac{\tilde{\vp}_i}{z_i}\right| \ll 1.\label{phidotini}
\end{align}
Considering that the scalar field to cosmic fluid energy densities ratio $\rho_\vp(z)/\rho_a(z)$, with $\rho_\vp(z)$ and $\rho_a(z)$ as given by the ``$z$-version'' of \eqref{energydensity} and \eqref{friedeq}, respectively, turns out to be
\begin{align}
\frac{\rho_\vp(z)}{\rho_a(z)} &= \frac{1}{3}\left(1 + z\right)^2\tilde{\vp}^{\prime}(z)^2 - 4\left(1 + z\right)\left[\beta - 4\,\gamma\,\chi(z) - 24\,\lambda\,\chi(z)^2\right]\tilde{\vp}(z)\tilde{\vp}^{\prime}(z)\nonumber\\
&+ 2\left[\beta + 8\,\lambda\,\chi(z)^2\right]\tilde{\varphi}(z)^{2},
\end{align}
one may foresee that the initial conditions \eqref{phiini} and \eqref{phidotini} are indeed compatible with the previous assumption that $\rho_\vp(z) \ll \rho_a(z)$, so the scalar field remains cosmologically subdominant, as required by observations. Actually, the fact of the matter is that a field value $\tilde{\vp} \sim 1$ constitutes a Planckian regime which is not compatible with an EFT description because such a scenario most probably falls into the so-called ``swampland'' of quantum gravity (see \cite{Palti:2019pca} and references therein); all in all, $\tilde{\vp}_i \ll 1$ should \textit{not} be considered fine tuning but the natural choice for the problem at hand.

\newpage

\begin{figure}[H]
\centering
\includegraphics[width=\textwidth]{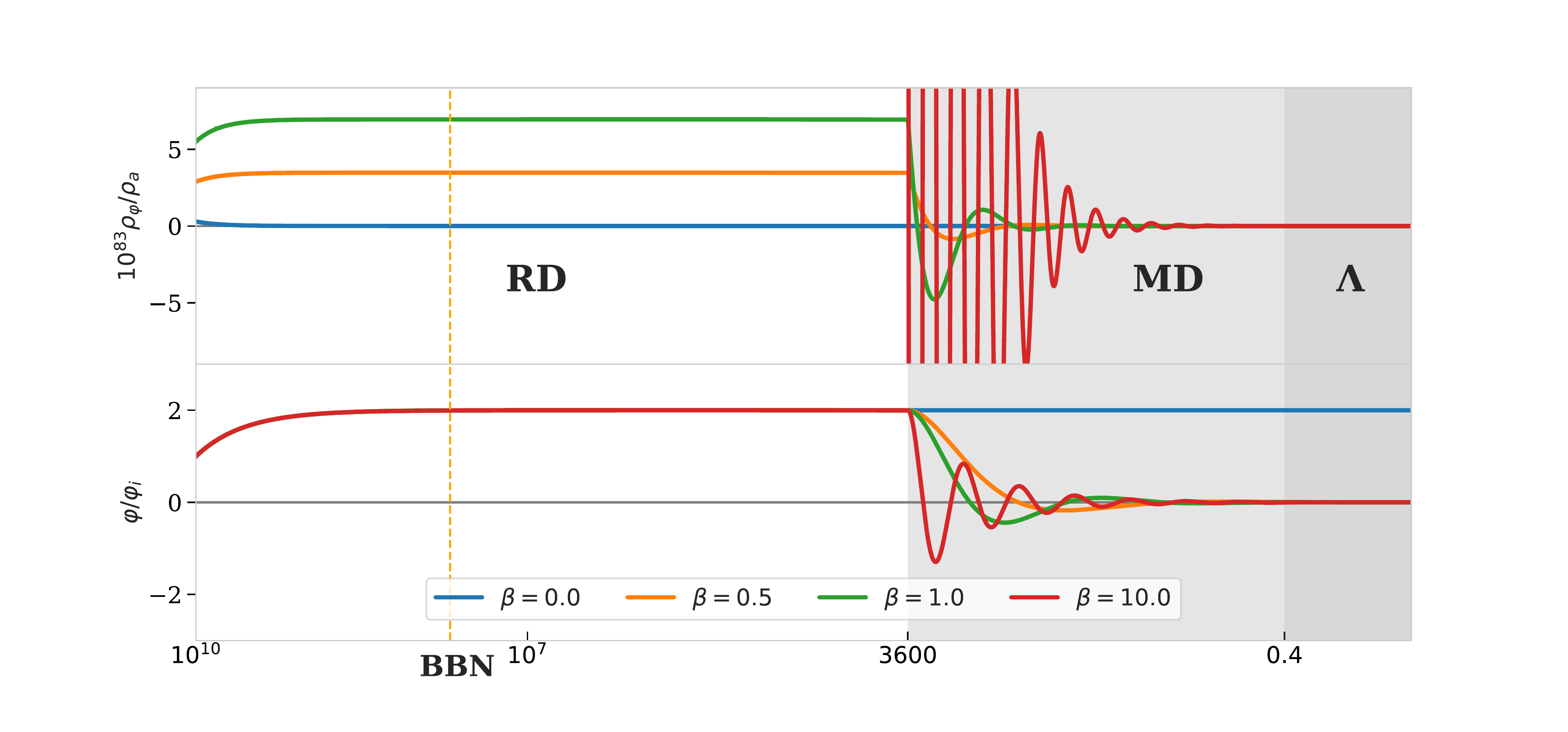}
\caption{{\it Top panel}: Effective energy density $\rho_\vp$ relative to the energy density of the cosmic fluid $\rho_a$. {\it Bottom panel}: Scalar field value relative to its initial value fixed at $z_i = 10^{10}$. The values of the coupling constants are taken to be $\gamma = 1$ and $\lambda = 48/175$, so they saturate the inequality in \eqref{constcoup}.}
\label{plot1}
\end{figure}

In Figure \ref{plot1} below we show the evolution of both the dimensionless scalar field $\varphi/\varphi_i$ and the dimensionless ratio $\rho_\vp/\rho_a$ for $z < z_i$ for different values of $\beta$ and fixed values of $\gamma$ and $\lambda$.
As previously noted, the contributions from higher-order curvature terms do become relevant during the very early stage of the universe. We may explicitly observe this, for high redshift $z > z_i$, in Figure \ref{plot2} below.
\begin{figure}[H]
\centering
\includegraphics[width=\textwidth]{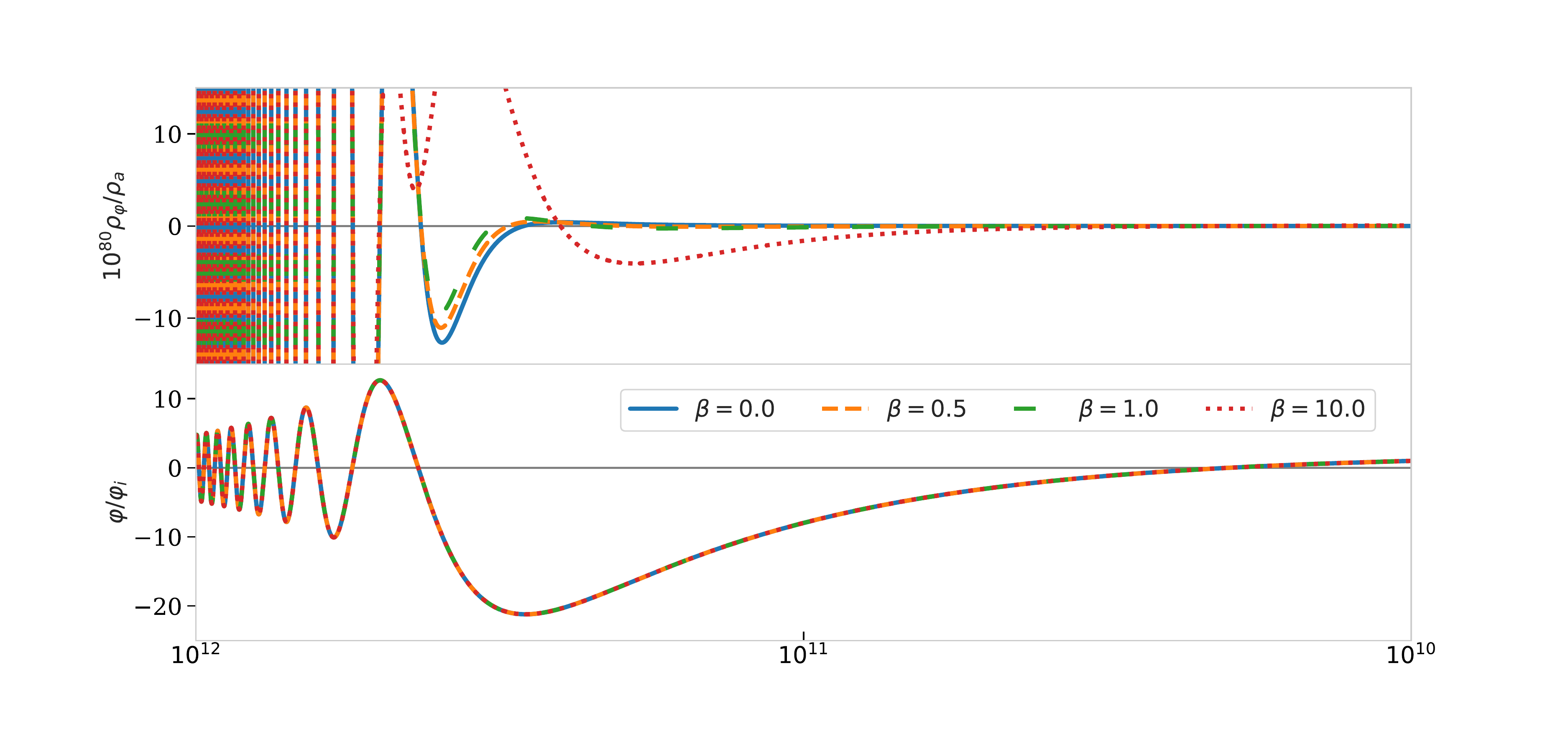}
\caption{Relative effective density and scalar field value for high redshift. The values of the coupling constants are the same as those of Figure \ref{plot1}.}
\label{plot2}
\end{figure}
We confirm that the solution is \textit{strongly} consistent with our initial assumption, as actually $\rho_\vp(z) \lll \rho_a(z)$ for the whole range of numerical integration which goes from $z = 0$ to ${z = 10^{12}}$. Moreover, $\vp$ is basically a constant throughout most of that range with the exception of cosmological ``phase transitions'' redshifts and very early and late times.

During early times, or high redshift, $\meff^2$ dominates over Hubble friction within the scalar field equation \eqref{SEM2}. However, as we ``move'' forward in time, $\meff^2$ decays much faster than the Hubble friction which rapidly takes over, so it is expected that the scalar field freezes to a constant way before entering MD era. To confirm that this is indeed the case, we may estimate the time at which the potential and the Hubble friction become comparable, since afterwards it should only take a couple of Hubble times for $\dot{\vp}$ to effectively vanish. Explicitly, using \eqref{meffRD}, the condition $m_\text{eff,R} \lesssim H$ is equivalent to $\sqrt{24\,\gamma\,\chi + 192\,\lambda\,\chi^2} \lesssim 1$, which by taking $\gamma = 1$, $\lambda = 48/175$ and $L = 10$ km., along with the fact that during RD $H(z) = H_0\left(1 + z\right)^2$ with $H_0 \simeq 10^{-43}$ GeV, comes to $z \approx 1.97 \times 10^{11}$. For even higher redshift values the relative scalar field and the relative energy density oscillate with ever increasing frequency as can be seen in Figure \ref{plot2}. We also explicity note that within this regime the relative scalar field is completely insensitive to the value of $\beta$ as the Ricci scalar identically vanishes, while the relative density does marginally depend on such a constant even though all the curves, for high enough $z$, eventually converge.  

On the other hand, by the time the MD era begins, the Ricci scalar stops being trivial, and in fact it entirely determines the relative scalar and energy density evolution because the higher-order operators become irrelevant considering that $\chi \sim 10^{-36} \lll 1$ when $z = 3600$. Looking at Figure \ref{plot1} we observe that the strict $\beta = 0$ case is actually problematic because the relative field stays frozen at its RD value and this goes against the premise of having an unscalarized $\vp_0 = 0$ asymptotic solution, which is actually mandatory, in view of weak field constraints. To the contrary, the same figure also shows that for any $\beta \neq 0$ scenario, the desired behavior of an asymptotically vanishing scalar field is attained. Moreover, as mentioned in \cite{Antoniou:2020nax}, $\vp$ oscillations may in principle affect the formation of large scale structures in the universe through the redefinition of (an effective) Newton's constant $G_\text{eff} \equiv G/\left(1 - 2\,\beta\,\tilde{\vp}^2\right)$ with $\tilde{\vp} \equiv \vp/\sqrt{2}\mp$ as before. However, it is clear that $\tilde{\vp}^2 \lll 1$ and, since $\beta \sim \mathcal{O}(1)$ number, $G_\text{eff} \approx G$ for all purposes, so GR truly becomes a trustworthy cosmological attractor. As it was expected, the scalar field profile in SECG exhibits a manifest deviation from its quadratic counterpart \cite{Antoniou:2020nax} for very high cosmological redshifts. This is depicted in Figure \ref{plot3} below.
\begin{figure}[H]
\centering
\includegraphics[width=\textwidth]{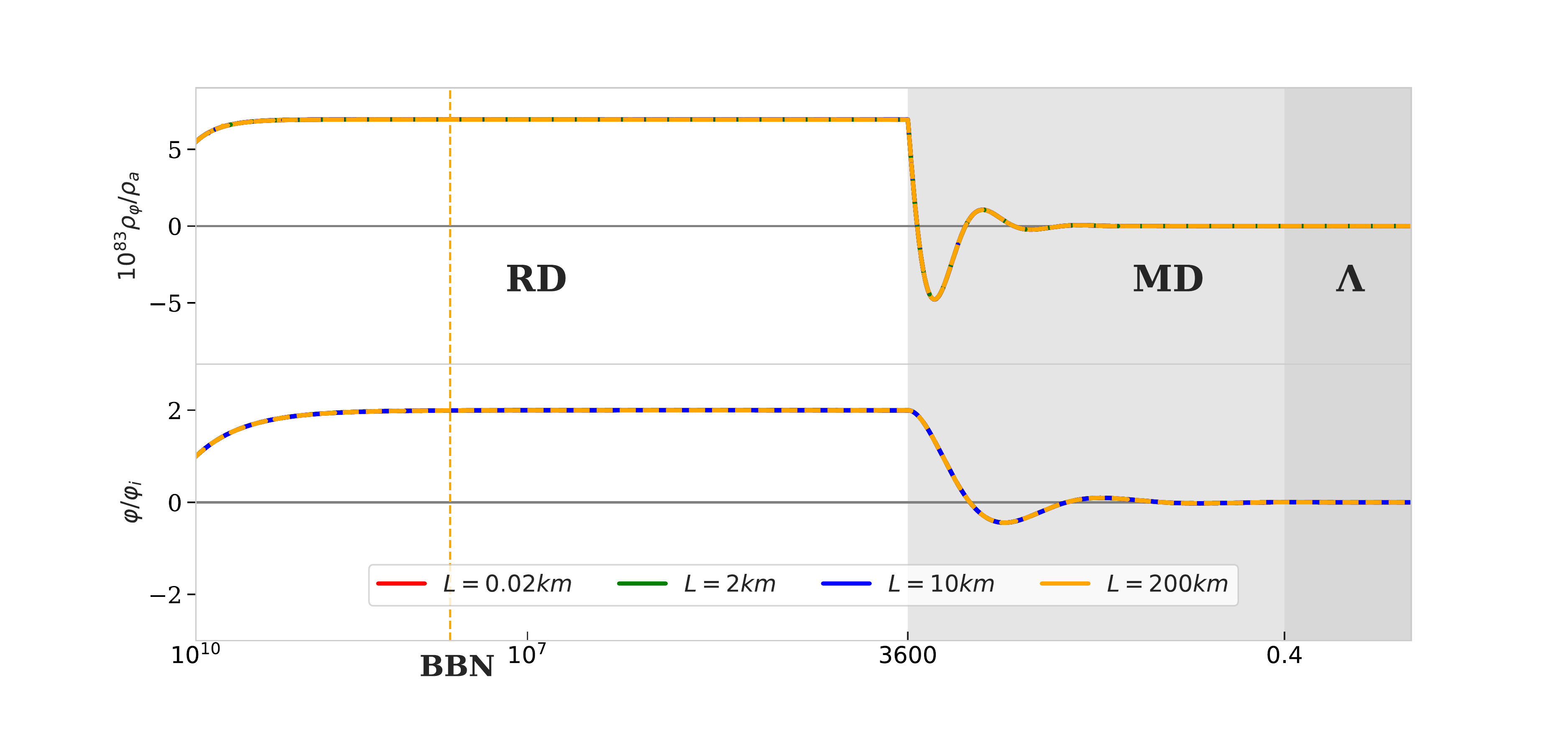}
\includegraphics[width=\textwidth]{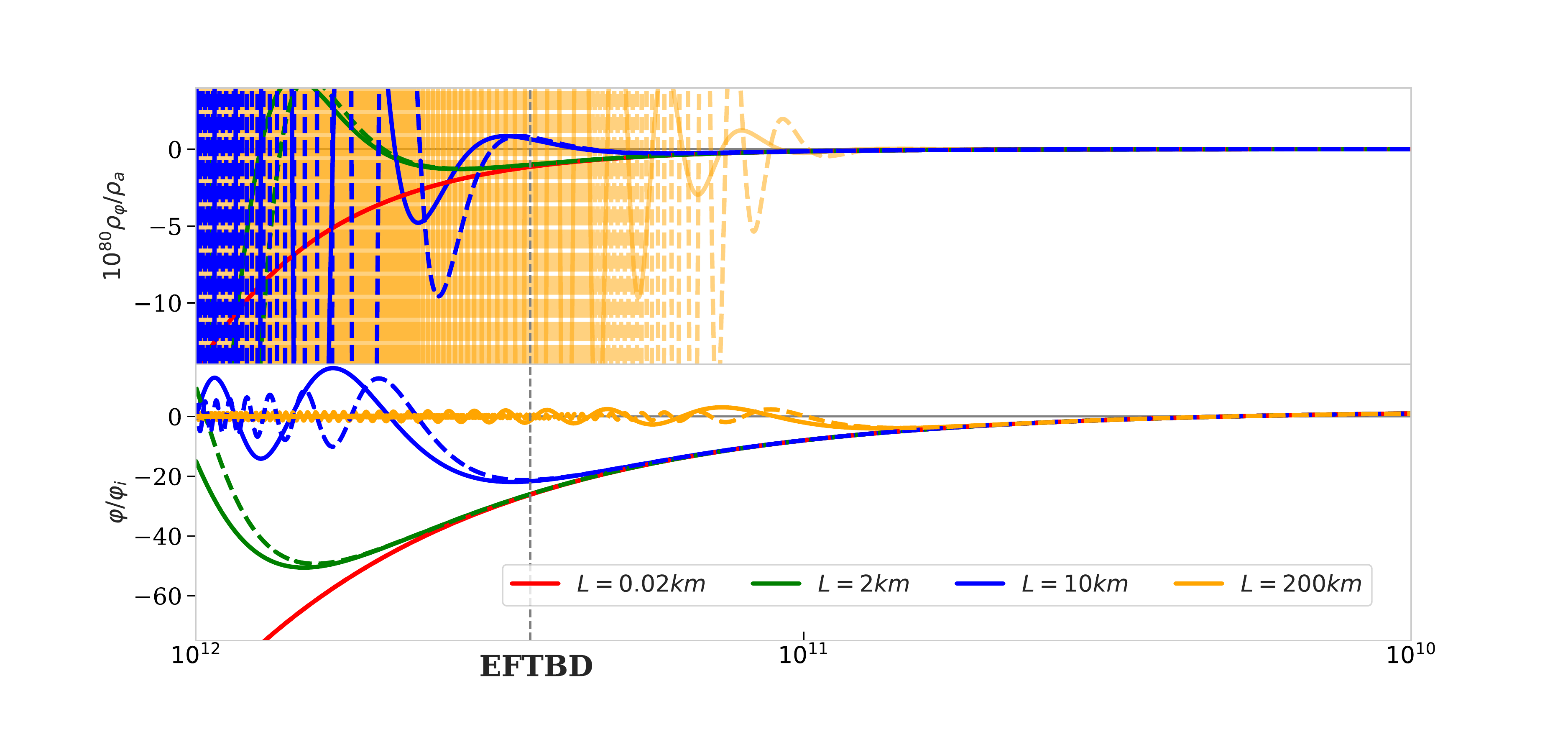}
\caption{Relative effective density and scalar field. The upper panel shows their value for $z < z_i$ while the lower one does so for $z > z_i$. Each color represents a theory with a specific value of the ``astrophysical length'' $L$. The continuous and dashed curves represent the profile stemming from the quadratic theory\cite{Antoniou:2020nax} and SECG, respectively. The values of the coupling constants are taken to be $\beta = 1$, $\gamma = 1$, and $\lambda = 48/175$.}
\label{plot3}
\end{figure}
In Figure \ref{plot3} each color stands for the profiles obtained within both the quadratic theory\cite{Antoniou:2020nax} (continuous curves) and SECG (dashed curves) for different values of the astrophysical length ${L = M^{-1}}$. We observe that the scalar field equation in SECG only gets a small correction (as it should) with respect to the Gauss-Bonnet quadratic theory during radiation domination up to a specific redshift value where both terms become ``competitive''. From \eqref{qr} we see that this roughly happens when $\gamma \approx 8\,\lambda\,\chi$ which implies, for the $L = 10$ km. case, a redshift value $z \approx 3.7 \times 10^{11}$, which lies before BBN epoch. Such a redshift value marks the ``breakdown'' of the EFT expansion (EFTBD), in the sense that perturbativity is lost and we should \textit{not} trust the naive model anymore. In other words, in reality, any behavior of the system beyond the EFTBD point should not be taken seriously as it does not represent sensible perturbative physics because the system becomes strongly-coupled.\footnote{As it is well-known, one possibility of rendering the theory perturbative again is that new degrees of freedom enter the game as it is further discussed in Appendix \ref{eftsnut}.} Effectively, by the time the cubic contribution becomes relevant, new non-trivial physics in the form of an ultraviolet (UV) completion, like inflation, necessarily becomes a much better description of the primordial cosmic plot. Interestingly, by the EFTBD time the size of the universe is about four kilometers. It can also be visually checked that in general, given that the redshift corresponding to the EFTBD point scales as $z \propto L^{-1/2}$, the larger the astrophysical length $L$, the smaller the redshift $z$ where the continuous and dashed curves formally ``separate''. This implies that the bigger the scale $M$, the further back in time the EFT is appropriate to describe the phenomenon of scalarization within a cosmological setting.

\section{Discussion and Conclusions}\label{discandconc}

In this paper we have followed the principles of EFT in order to address the issue of black hole scalarization and its compatibility with standard cosmic evolution in higher-order theories of gravity, using what we have dubbed ``Scalar-Einstenian Cubic Gravity'' as a well-motivated, healthy toy model. Within the process, we have learned several lessons concerning the naive approach one may have to such a seemingly straightforward model-building exercise. For example, it became obvious by using ``naturalness'' arguments, the complete necessity of introducing a ``foreign'' new scale in the problem, which is actually associated with the physical extension of the would-be scalarized compact object. Interestingly enough, such a procedure also improves the phenomenological prospects for scalarization in the sense that, without a new scale, the mass of the black holes that may become scalarized would be of order $10^{-37}$ solar masses, so they clearly lie way out of any observational scope. Moreover, we have confirmed the fact that the relative signs of the dimensionless parameters of the model play a crucial role within the process of spacetime scalarization as they actually determine whether or not scalarized solutions may emerge from dynamics.  

In the context of tackling the phenomenon of scalarization through the construction of a model in a systematic manner we conclude that the claim for the existence of a catastrophic instability during inflation \cite{Anson:2019uto}, although classically true, is not really justified from an EFT perspective. This is because there is \textit{no} symmetry reason for discarding the marginal operator $\vp^2 R$, which shall be quantum-mechanically generated anyway, and will actually lead the dynamics whenever the theory is weakly-coupled. Moreover, picking the right sign for such an operator in the theory is enough to render inflation safe from the tachyonic instability while still scalarizing compact objects through the irrelevant operator $\vp^2\G$, hence \textit{scalarization remains compatible with the inflationary paradigm.}

\newpage

Next we addressed the issue of the compatibility of scalarization and BBC evolution. To do so, we followed \cite{Antoniou:2020nax} and integrated the scalar field equation to find that, indeed under very sensible assumptions for the initial conditions of the system, \textit{the theory admits GR as a cosmological attractor.} The departure from GR does become significant for high redshift $\left(z \sim 10^{11}\right)$, way before BBN, where the relative scalar field and energy density deviate from its quasi-constant behavior. The departure, however, between SECG theory and the quadratic one, occurs around the same high redshift and this roughly defines the breakdown of the weakly-coupled EFT description, implying the necessity of a UV completion. The understanding of the interplay between the different scales in the game is further elucidated in Appendix \ref{eftsnut}.

Finally, let us note that it is quite remarkable that by using a criterion based on the existence of an unstable mode for a scalar field fluctuation living on a Schwarzschild background we may get important information about the appropriate EFT that captures the phenomenon of black hole scalarization. In short, such an unstable mode signals the instability of the underlying geometry as the curvature surpasses a specific threshold, and its tendency to consequently develop ``scalar hair'', i.e, a non-zero scalar profile. By imposing a sensible reality condition for the would-be scalarized black hole mass we were able to find a non-trivial constraint between otherwise completely independent ``Wilson coefficients'' of the EFT expansion. It would be nice to find, through the application of some other physical requirements, more relations between the EFT coefficients so that we would be able to further constrain the parameter space of the model.

For example, if we abandon the goal of scalarizing astrophysical objects by entertaining the idea of dealing with more fundamental ``primordial'' black holes, we may consider the free theory of scalar fluctuations on a (quasi)-dS inflationary spacetime, whose analytic solutions are well-known. As detailed in Appendix \ref{spertds}, the relevant quantity for the scalar fluctuation dynamics in a dS background is the ``index'' ${\nu^2 \equiv \frac{9}{4} - \frac{\meff^2}{H^2}}$, which actually needs \textit{not} to be necessarily positive. The whole ``cosmological collider physics'' programme of Arkani-Hamed and Maldacena \cite{Arkani-Hamed:2015bza}, which emphasizes the fact that inflation pertains the highest energy \textit{observable} natural process, is actually quite simple for scalar fields; particles of mass $m \sim H$ during inflation may leave observable characteristic imprints, so they become ``fossils'' of a fundamental high-energy physics era, depending on the actual value of the $\nu$-index. Then, as our scalar ``particle'' has a zeroth-order $\nu$-index given by $\nu^2 \simeq \frac{9}{4} - 12\,\beta$, the imprint of such a particle would be linked to the value of the EFT parameter $\beta$, which we only demand to be positive (so BBC does \textit{not} get scalarized) but whose magnitude is otherwise undetermined. This opens up the possibility of eventually measuring the actual value of the $\beta$ parameter. More kinematical and dynamical conditions among EFT parameters are waiting to be explored and it is our aim to do so in the near future.

Let us close by stating that it would be great to derive that $\beta$ ultimately \textit{needs} to be positive in order for the EFT to be compatible with a quantum gravity UV-completion, using the so-called ``positivity constraints'' programme \cite{Adams:2006sv}. Moreover, we want to understand under what circumstances the shift symmetry is only spoiled by gravitational interactions. We are already working in a realistic particle physics model that truly embeds scalarization within the framework of spontaneous symmetry breaking, and we shall report our findings elsewhere \cite{inprep}. 

\section*{Acknowledgements}

 CE is supported by PAI Grant N\textsuperscript{\underline{o}} 77190046, by FONDECYT Grant N\textsuperscript{\underline{o}} 11221063  and Universidad Central de Chile through Proyecto Interno CIP2020039. CE would like to thank the
Department of Physics at the National Technical University of Athens (Greece) and Centro de Estudios Cient\'ificos at Valdivia (Chile) for warm hospitality. SR acknowledges support from Fondecyt Postdoctorado project N\textsuperscript{\underline{o}} 3190554.

\begin{appendices}
\section{EFT's in a Nutshell.}\label{eftsnut}
\renewcommand{\theequation}{A.\arabic{equation}}
\setcounter{equation}{0}
Say the final theory of some physical system may be written in terms of two fields $\phi$ and $\sigma$, meaning the UV complete theory has a Lagrangian of the form
\begin{align}
    \L[\phi,\sigma] = \L[\phi] + \L[\sigma] + \L_\text{int}[\phi,\sigma].
\end{align}
We will assume that $\phi$ is a \textit{light} field while $\sigma$ is a \textit{heavy} field, meaning $m_\phi < \Lambda$ and $M_\sigma > \Lambda$, where $m_\phi$ and $M_\sigma$ are the masses of $\phi$ and $\sigma$, respectively, and $\Lambda$ is the cut-off of the EFT. To derive the EFT, encapsulated in the effective action $S_\text{eff}$, we use the so-called \textit{Wilsonian approach} formally defined through the path integral
\begin{align}
    e^{iS_\text{eff}[\phi]} = \int\mathscr{D}\sigma\,e^{iS[\phi,\sigma]}.
\end{align}
As it is well-known, at tree-level, performing such an integral is equivalent to using the field equations to eliminate the heavy field. The complete path integral includes, however, loop corrections involving the heavy fields, implying that the effective action is, in principle, non-local. However, at low energies, meaning $E \ll \Lambda \leq M_\sigma$, one may get an approximately local EFT action with a systematic expansion of the form
\begin{align}
    \L_\text{eff}[\phi] = \L[\phi] + \sum_ic_i(g)\,\frac{\mathcal{O}_i[\phi]}{(M_\sigma)^{\Delta_i - 4}},
\end{align}
where the $c_i = c_i(g)$ are dimensionless constants which are expected to be of order one. It is important to emphasize that the meaning of $c_i$ being order one numbers is somehow \textit{not} universal. In this paper we declare that any number $c_i \sim \left\{10^{-2},10^2\right\}$ should be considered an order one number (basically, not super small, not super big). Anyhow, these numbers depend on the couplings $g$ of the UV theory, while the $\mathcal{O}_i$ are local operators of mass dimension $\Delta_i$. The procedure described above will generate all operators compatible with the symmetries of the UV theory, implying that the absence of any of such terms, or an anomalously small corresponding coefficient, should be understood as fine-tuning (an important exception to this statement is the case of a so-called \textit{accidental symmetry}, which however, is \textit{not} relevant for our discussion, so we do not consider it).

We may take the point of view that GR and any low-energy field description encapsulate the relevant degrees of freedom for describing the physics way below the Planck scale $\mp$. Therefore, for a higher-order EFT of gravity such as the one we deal with in this paper, it is natural to take $\mp$ as the cut-off and so we do. The introduction of the scale $M$ however, does become necessary in order to ``naturalize'' otherwise finely-tuned coefficients of the EFT, as emphasized elsewhere within the article. In principle, one may think that the operators containing such a scale are the remains after integrating out dynamical degrees of freedom of mass $M$. This implies that it is natural to expect the hierarchy $\meff^2 \ll M^2 \ll \mp^2$. However, the perturbative condition $H^2 \ll M^2$ is not fundamental, but a restriction of the weakly-coupled EFT description.

\section{Scalar Perturbation in a de Sitter Background.}\label{spertds}
\renewcommand{\theequation}{B.\arabic{equation}}
\setcounter{equation}{0}
Let us consider dS spacetime in conformal coordinates $dt = a(\eta)d\eta$ with line element  
\begin{align}
ds^2 = \frac{-d\eta^2 + d\boldsymbol{x}^2}{H^2\eta^2}.
\end{align}
In such coordinates the Klein-Gordon equation for the scalar fluctuation, $\left(\Box - \meff^2\right)\delta\vp = 0$, reads
\begin{align}
    \delta\vp_{\boldsymbol{k}}'' - \frac{2}{\eta}\delta\vp_{\boldsymbol{k}}' + k^2\delta\vp_{\boldsymbol{k}} + \frac{\meff^2}{H^2\eta^2}\delta\vp_{\boldsymbol{k}} = 0,
\end{align}
where $\meff^2 \equiv 12(\beta  - 2\gamma\chi - 8\lambda\chi^2)H^2$, $\chi \equiv (H/M)^2$, and $\delta\vp_{\boldsymbol{k}}' \equiv \frac{d\delta\vp_{\boldsymbol{k}}}{d\eta}$ within this appendix. Defining the field $\psi \equiv a\,\delta\vp$, the above equation may be put in a ``Mukhanov-Sasaki'' form, meaning
\begin{align}
    \psi_{\boldsymbol{k}}'' + \left(k^2 - \frac{\nu^2 - \frac{1}{4}}{\eta^2}\right)\psi_{\boldsymbol{k}} = 0,\label{MSeqn}
\end{align}
where $\nu \equiv \sqrt{\frac{9}{4} - \frac{\meff^2}{H^2}}$. The solution to \eqref{MSeqn} is well-known. It is given by
\begin{align}
    \psi_{\boldsymbol{k}} &= -ie^{i\left(\nu + \frac{1}{2}\right)\frac{\pi}{2}}\frac{\sqrt{\pi}}{2}H(-\eta)^{3/2}H_\nu^{(1)}(-k\eta)\quad &\text{if} \quad \frac{\meff^2}{H^2} \leq \frac{9}{4},\label{sol1}\\
    \psi_{\boldsymbol{k}} &= -ie^{-\frac{\pi}{2}\tilde{\nu} + i\frac{\pi}{4}}\frac{\sqrt{\pi}}{2}H(-\eta)^{3/2}H_{i\tilde{\nu}}^{(1)}(-k\eta)\quad &\text{if} \quad \frac{\meff^2}{H^2} > \frac{9}{4},\label{sol2}
\end{align}
where $\tilde{\nu} \equiv \sqrt{\frac{\meff^2}{H^2} - \frac{9}{4}}$, and usual Bunch-Davies initial conditions have been imposed. It is useful to consider the behavior of the mode functions after horizon exit, $k\eta \to 0$. One finds that for $\meff^2/H^2 \leq 9/4$,
\begin{align}
    \psi_{\boldsymbol{k}} &\to -e^{i\left(\nu + \frac{1}{2}\right)\frac{\pi}{2}}\frac{2^{\nu - 1}}{\sqrt{\pi}}\Gamma(\nu)\frac{H}{k^{\nu}}\left(-\eta\right)^{-\nu + \frac{3}{2}},\quad &0 < \nu \leq \frac{3}{2},\nonumber\\
    \psi_{\boldsymbol{k}} &\to e^{i\frac{\pi}{4}}\frac{1}{\sqrt{\pi}}H\left(-\eta\right)^{3/2}\ln\left(-k\eta\right), &\nu = 0,\label{lim1}
\end{align}
and for $\meff^2/H^2 > 9/4$,
\begin{align}
    \psi_{\boldsymbol{k}} \to -ie^{-\frac{\pi}{2}\tilde{\nu} + i\frac{\pi}{4}}\frac{\sqrt{\pi}}{2}H\left(-\eta\right)^{3/2}\left[\frac{1}{\Gamma\left(i\tilde{\nu} + 1\right)}\left(\frac{-k\eta}{2}\right)^{i\tilde{\nu}} - i\frac{\Gamma\left(i\tilde{\nu}\right)}{\pi}\left(\frac{-k\eta}{2}\right)^{-i\tilde{\nu}}\right].\label{lim2}
\end{align}
Recalling that during inflation $H \simeq 10^{13}$ GeV, it is crucial to impose a perturbative hierarchy ${H^2 \ll M^2 \ll \mp^2}$, so that ${\chi \ll 1}$, and it is then safe to approximate $\meff^2 \simeq 12\beta H^2$. The latter approximation implies that $\nu^2 \simeq \frac{9}{4} - 12\beta$, so that \eqref{sol1} and \eqref{sol2} are appropriate whenever $\beta \lesssim \frac{3}{16}$ and $\beta > \frac{3}{16}$, respectively. Notice that by taking, for definiteness, $M \in \left[10^{14}, 10^{17}\right]$ GeV, implies we are dealing with a would-be scalarized object of physical length ${L \in \left[1.23\times 10^{5},123\right]\,\ell_\text{Pl}}$, with $\ell_\text{Pl} = 1.61 \times 10^{-35}$ m standing for the famous Planck length. 

We see that in the $k\eta \to 0$ limit, \eqref{lim1} contains a decay factor $\left(-\eta\right)^{-\nu + 3/2}$, while \eqref{lim2} contains the decay factor $\left(-\eta\right)^{3/2}$ and an oscillation factor of the form $\eta^{\pm i\tilde{\nu}}$. Both decay factors indicate the well-known fact that massive field perturbations eventually die off. In the context of inflation, information about the mass spectrum is encoded in the power of the momentum ratio in the squeezed-limit bispectrum (see \cite{Meerburg:2016zdz} for more details). In short, the squeezed-limit of the bispectrum may be considered as a ``particle detector'' for massive particles. Importantly, it is possible to show that for large masses, meaning $\meff > \mathcal{O}\left(H\right)$, the signal acquires a ``Boltzmann suppression'' $e^{-\pi\tilde{\nu}}$ since too massive particles cannot be thermally produced during inflation, so this factor limits the mass range for such a particle detector. Therefore one needs to focus on particles whose masses are not too much heavier than $H$. This is exactly our case at hand, because $\meff \simeq \sqrt{12\beta}H$, with $\beta \sim \mathcal{O}\left(1\right)$ number, as expected from an EFT perspective.

\newpage

\end{appendices}

\bibliographystyle{JHEP}
\bibliography{refs}

\providecommand{\href}[2]{#2}\begingroup\raggedright\begin{thebibliography}{10}

\bibitem{Abbott1}
{\scshape LIGO Scientific, Virgo} collaboration, \emph{{Observation of
  Gravitational Waves from a Binary Black Hole Merger}},
  \href{https://doi.org/10.1103/PhysRevLett.116.061102}{\emph{Phys. Rev. Lett.}
  {\bfseries 116} (2016) 061102}
  [\href{https://arxiv.org/abs/1602.03837}{{\ttfamily 1602.03837}}].

\bibitem{Abbott2}
{\scshape LIGO Scientific, Virgo} collaboration, \emph{{GW170817: Observation
  of Gravitational Waves from a Binary Neutron Star Inspiral}},
  \href{https://doi.org/10.1103/PhysRevLett.119.161101}{\emph{Phys. Rev. Lett.}
  {\bfseries 119} (2017) 161101}
  [\href{https://arxiv.org/abs/1710.05832}{{\ttfamily 1710.05832}}].

\bibitem{Abbott3}
{\scshape LIGO Scientific, Virgo, Fermi-GBM, INTEGRAL} collaboration,
  \emph{{Gravitational Waves and Gamma-rays from a Binary Neutron Star Merger:
  GW170817 and GRB 170817A}},
  \href{https://doi.org/10.3847/2041-8213/aa920c}{\emph{Astrophys. J. Lett.}
  {\bfseries 848} (2017) L13}
  [\href{https://arxiv.org/abs/1710.05834}{{\ttfamily 1710.05834}}].

\bibitem{def1}
T.~Damour and K.~Nordtvedt, \emph{General relativity as a cosmological
  attractor of tensor-scalar theories},
  \href{https://doi.org/10.1103/PhysRevLett.70.2217}{\emph{Phys. Rev. Lett.}
  {\bfseries 70} (1993) 2217}.

\bibitem{def2}
T.~Damour and G.~Esposito-Far\`ese, \emph{Nonperturbative strong-field effects
  in tensor-scalar theories of gravitation},
  \href{https://doi.org/10.1103/PhysRevLett.70.2220}{\emph{Phys. Rev. Lett.}
  {\bfseries 70} (1993) 2220}.

\bibitem{donevaprl}
D.D.~Doneva and S.S.~Yazadjiev, \emph{{New Gauss-Bonnet Black Holes with
  Curvature-Induced Scalarization in Extended Scalar-Tensor Theories}},
  \href{https://doi.org/10.1103/PhysRevLett.120.131103}{\emph{Phys. Rev. Lett.}
  {\bfseries 120} (2018) 131103}
  [\href{https://arxiv.org/abs/1711.01187}{{\ttfamily 1711.01187}}].

\bibitem{donevaprl2}
H.O.~Silva, J.~Sakstein, L.~Gualtieri, T.P.~Sotiriou and E.~Berti,
  \emph{Spontaneous scalarization of black holes and compact stars from a
  gauss-bonnet coupling},
  \href{https://doi.org/10.1103/PhysRevLett.120.131104}{\emph{Phys. Rev. Lett.}
  {\bfseries 120} (2018) 131104}.

\bibitem{stelle}
K.S.~Stelle, \emph{{Renormalization of Higher Derivative Quantum Gravity}},
  \href{https://doi.org/10.1103/PhysRevD.16.953}{\emph{Phys. Rev. D} {\bfseries
  16} (1977) 953}.

\bibitem{stewart}
S.~Mignemi and N.R.~Stewart, \emph{{Charged black holes in effective string
  theory}}, \href{https://doi.org/10.1103/PhysRevD.47.5259}{\emph{Phys. Rev. D}
  {\bfseries 47} (1993) 5259}
  [\href{https://arxiv.org/abs/hep-th/9212146}{{\ttfamily hep-th/9212146}}].

\bibitem{kanti}
P.~Kanti, N.E.~Mavromatos, J.~Rizos, K.~Tamvakis and E.~Winstanley,
  \emph{{Dilatonic black holes in higher curvature string gravity}},
  \href{https://doi.org/10.1103/PhysRevD.54.5049}{\emph{Phys. Rev. D}
  {\bfseries 54} (1996) 5049}
  [\href{https://arxiv.org/abs/hep-th/9511071}{{\ttfamily hep-th/9511071}}].

\bibitem{torii2}
T.~Torii, H.~Yajima and K.-i.~Maeda, \emph{{Dilatonic black holes with
  Gauss-Bonnet term}},
  \href{https://doi.org/10.1103/PhysRevD.55.739}{\emph{Phys. Rev. D} {\bfseries
  55} (1997) 739} [\href{https://arxiv.org/abs/gr-qc/9606034}{{\ttfamily
  gr-qc/9606034}}].

\bibitem{yunes}
D.~Ayzenberg and N.~Yunes, \emph{{Slowly-Rotating Black Holes in
  Einstein-Dilaton-Gauss-Bonnet Gravity: Quadratic Order in Spin Solutions}},
  \href{https://doi.org/10.1103/PhysRevD.90.044066}{\emph{Phys. Rev. D}
  {\bfseries 90} (2014) 044066}
  [\href{https://arxiv.org/abs/1405.2133}{{\ttfamily 1405.2133}}].

\bibitem{kleihaus}
B.~Kleihaus, J.~Kunz and E.~Radu, \emph{{Rotating Black Holes in Dilatonic
  Einstein-Gauss-Bonnet Theory}},
  \href{https://doi.org/10.1103/PhysRevLett.106.151104}{\emph{Phys. Rev. Lett.}
  {\bfseries 106} (2011) 151104}
  [\href{https://arxiv.org/abs/1101.2868}{{\ttfamily 1101.2868}}].

\bibitem{doneva33}
D.D.~Doneva and S.S.~Yazadjiev, \emph{{Neutron star solutions with curvature
  induced scalarization in the extended Gauss-Bonnet scalar-tensor theories}},
  \href{https://doi.org/10.1088/1475-7516/2018/04/011}{\emph{JCAP} {\bfseries
  04} (2018) 011} [\href{https://arxiv.org/abs/1712.03715}{{\ttfamily
  1712.03715}}].

\bibitem{bakopoulos1}
G.~Antoniou, A.~Bakopoulos and P.~Kanti, \emph{Evasion of no-hair theorems and
  novel black-hole solutions in gauss-bonnet theories},
  \href{https://doi.org/10.1103/physrevlett.120.131102}{\emph{Physical Review
  Letters} {\bfseries 120} (2018) }.

\bibitem{bakopoulos2}
G.~Antoniou, A.~Bakopoulos and P.~Kanti, \emph{Black-hole solutions with scalar
  hair in einstein-scalar-gauss-bonnet theories},
  \href{https://doi.org/10.1103/physrevd.97.084037}{\emph{Physical Review D}
  {\bfseries 97} (2018) }.

\bibitem{Doneva2}
D.D.~Doneva, S.~Kiorpelidi, P.G.~Nedkova, E.~Papantonopoulos and
  S.S.~Yazadjiev, \emph{Charged gauss-bonnet black holes with curvature induced
  scalarization in the extended scalar-tensor theories},
  \href{https://doi.org/10.1103/physrevd.98.104056}{\emph{Physical Review D}
  {\bfseries 98} (2018) }.

\bibitem{herdeiro1}
C.A.~Herdeiro, E.~Radu, N.~Sanchis-Gual and J.A.~Font, \emph{Spontaneous
  scalarization of charged black holes},
  \href{https://doi.org/10.1103/physrevlett.121.101102}{\emph{Physical Review
  Letters} {\bfseries 121} (2018) }.

\bibitem{herdeiro2}
P.G.S.~Fernandes, C.A.R.~Herdeiro, A.M.~Pombo, E.~Radu and N.~Sanchis-Gual,
  \emph{Spontaneous scalarisation of charged black holes: coupling dependence
  and dynamical features},
  \href{https://doi.org/10.1088/1361-6382/ab23a1}{\emph{Classical and Quantum
  Gravity} {\bfseries 36} (2019) 134002}.

\bibitem{herdeiro3}
C.A.~Herdeiro and E.~Radu, \emph{Black hole scalarization from the breakdown of
  scale invariance},
  \href{https://doi.org/10.1103/physrevd.99.084039}{\emph{Physical Review D}
  {\bfseries 99} (2019) }.

\bibitem{herdeiro4}
Y.~Brihaye, C.~Herdeiro and E.~Radu, \emph{The scalarised schwarzschild-nut
  spacetime},
  \href{https://doi.org/10.1016/j.physletb.2018.11.022}{\emph{Physics Letters
  B} {\bfseries 788} (2019) 295}.

\bibitem{Doneva3}
D.D.~Doneva, K.V.~Staykov, S.S.~Yazadjiev and R.Z.~Zheleva, \emph{Multiscalar
  gauss-bonnet gravity: Hairy black holes and scalarization},
  \href{https://doi.org/10.1103/physrevd.102.064042}{\emph{Physical Review D}
  {\bfseries 102} (2020) }.

\bibitem{astefanesei}
D.~Astefanesei, C.~Herdeiro, J.~Oliveira and E.~Radu, \emph{Higher dimensional
  black hole scalarization},
  \href{https://doi.org/10.1007/jhep09(2020)186}{\emph{Journal of High Energy
  Physics} {\bfseries 2020} (2020) }.

\bibitem{canate}
P.~Ca{\~n}ate and S.E.P.~Bergliaffa, \emph{Novel exact magnetic black hole
  solution in four-dimensional extended scalar-tensor-gauss-bonnet theory},
  \href{https://doi.org/10.1103/physrevd.102.104038}{\emph{Physical Review D}
  {\bfseries 102} (2020) }.

\bibitem{bakopoulos3}
A.~Bakopoulos, G.~Antoniou and P.~Kanti, \emph{Novel black-hole solutions in
  einstein-scalar-gauss-bonnet theories with a cosmological constant},
  \href{https://doi.org/10.1103/physrevd.99.064003}{\emph{Physical Review D}
  {\bfseries 99} (2019) }.

\bibitem{herdeiro5}
Y.~Brihaye, C.~Herdeiro and E.~Radu, \emph{Black hole spontaneous scalarisation
  with a positive cosmological constant},
  \href{https://doi.org/10.1016/j.physletb.2020.135269}{\emph{Physics Letters
  B} {\bfseries 802} (2020) 135269}.

\bibitem{bakopoulos4}
A.~Bakopoulos, P.~Kanti and N.~Pappas, \emph{Large and ultracompact
  gauss-bonnet black holes with a self-interacting scalar field},
  \href{https://doi.org/10.1103/physrevd.101.084059}{\emph{Physical Review D}
  {\bfseries 101} (2020) }.

\bibitem{lin}
K.~Lin, S.~Zhang, C.~Zhang, X.~Zhao, B.~Wang and A.~Wang, \emph{No static
  regular black holes in einstein-complex-scalar-gauss-bonnet gravity},
  \href{https://doi.org/10.1103/physrevd.102.024034}{\emph{Physical Review D}
  {\bfseries 102} (2020) }.

\bibitem{pap3}
H.~Guo, S.~Kiorpelidi, X.-M.~Kuang, E.~Papantonopoulos, B.~Wang and J.-P.~Wu,
  \emph{Spontaneous holographic scalarization of black holes in
  einstein-scalar-gauss-bonnet theories},
  \href{https://doi.org/10.1103/physrevd.102.084029}{\emph{Physical Review D}
  {\bfseries 102} (2020) }.

\bibitem{brihaye}
Y.~Brihaye, B.~Hartmann, N.~Pio~Aprile and J.~Urrestilla, \emph{Scalarization
  of asymptotically anti--de sitter black holes with applications to
  holographic phase transitions},
  \href{https://doi.org/10.1103/physrevd.101.124016}{\emph{Physical Review D}
  {\bfseries 101} (2020) }.

\bibitem{pap4}
Z.-Y.~Tang, B.~Wang, T.~Karakasis and E.~Papantonopoulos, \emph{{Curvature
  scalarization of black holes in f(R) gravity}},
  \href{https://doi.org/10.1103/PhysRevD.104.064017}{\emph{Phys. Rev. D}
  {\bfseries 104} (2021) 064017}
  [\href{https://arxiv.org/abs/2008.13318}{{\ttfamily 2008.13318}}].

\bibitem{kleihaus2}
L.G.~Collodel, B.~Kleihaus, J.~Kunz and E.~Berti, \emph{Spinning and excited
  black holes in einstein-scalar-gauss--bonnet theory},
  \href{https://doi.org/10.1088/1361-6382/ab74f9}{\emph{Classical and Quantum
  Gravity} {\bfseries 37} (2020) 075018}.

\bibitem{sotiriou4}
A.~Dima, E.~Barausse, N.~Franchini and T.P.~Sotiriou, \emph{Spin-induced black
  hole spontaneous scalarization},
  \href{https://doi.org/10.1103/physrevlett.125.231101}{\emph{Physical Review
  Letters} {\bfseries 125} (2020) }.

\bibitem{Doneva4}
D.D.~Doneva, L.G.~Collodel, C.J.~Kr{\"u}ger and S.S.~Yazadjiev,
  \emph{Spin-induced scalarization of kerr black holes with a massive scalar
  field}, \href{https://doi.org/10.1140/epjc/s10052-020-08765-3}{\emph{The
  European Physical Journal C} {\bfseries 80} (2020) }.

\bibitem{herdeiro6}
C.A.~Herdeiro, E.~Radu, H.O.~Silva, T.P.~Sotiriou and N.~Yunes,
  \emph{Spin-induced scalarized black holes},
  \href{https://doi.org/10.1103/physrevlett.126.011103}{\emph{Physical Review
  Letters} {\bfseries 126} (2021) }.

\bibitem{kleihaus3}
E.~Berti, L.G.~Collodel, B.~Kleihaus and J.~Kunz, \emph{Spin-induced black hole
  scalarization in einstein-scalar-gauss-bonnet theory},
  \href{https://doi.org/10.1103/physrevlett.126.011104}{\emph{Physical Review
  Letters} {\bfseries 126} (2021) }.

\bibitem{wang}
S.-J.~Zhang, B.~Wang, A.~Wang and J.F.~Saavedra, \emph{Object picture of scalar
  field perturbation on kerr black hole in scalar-einstein-gauss-bonnet
  theory}, \href{https://doi.org/10.1103/physrevd.102.124056}{\emph{Physical
  Review D} {\bfseries 102} (2020) }.

\bibitem{bueno1}
P.~Bueno and P.A.~Cano, \emph{{Einsteinian cubic gravity}},
  \href{https://doi.org/10.1103/PhysRevD.94.104005}{\emph{Phys. Rev.}
  {\bfseries D94} (2016) 104005}
  [\href{https://arxiv.org/abs/1607.06463}{{\ttfamily 1607.06463}}].

\bibitem{bueno2}
P.~Bueno and P.A.~Cano, \emph{Four-dimensional black holes in einsteinian cubic
  gravity}, \href{https://doi.org/10.1103/PhysRevD.94.124051}{\emph{Phys. Rev.
  D} {\bfseries 94} (2016) 124051}.

\bibitem{mann1}
R.A.~Hennigar and R.B.~Mann, \emph{Black holes in einsteinian cubic gravity},
  \href{https://doi.org/10.1103/PhysRevD.95.064055}{\emph{Phys. Rev. D}
  {\bfseries 95} (2017) 064055}.

\bibitem{mann2}
R.A.~Hennigar, D.~Kubiz\ifmmode~\check{n}\else \v{n}\fi{}\'ak and R.B.~Mann,
  \emph{Generalized quasitopological gravity},
  \href{https://doi.org/10.1103/PhysRevD.95.104042}{\emph{Phys. Rev. D}
  {\bfseries 95} (2017) 104042}.

\bibitem{edelstein1}
G.~Arciniega, J.D.~Edelstein and L.G.~Jaime, \emph{Towards geometric inflation:
  The cubic case},
  \href{https://doi.org/10.1016/j.physletb.2020.135272}{\emph{Physics Letters
  B} {\bfseries 802} (2020) 135272}.

\bibitem{cisternaqtg}
A.~Cisterna, N.~Grandi and J.~Oliva, \emph{On four-dimensional einsteinian
  gravity, quasitopological gravity, cosmology and black holes},
  \href{https://doi.org/10.1016/j.physletb.2020.135435}{\emph{Physics Letters
  B} {\bfseries 805} (2020) 135435}.

\bibitem{edelstein2}
G.~Arciniega, P.~Bueno, P.A.~Cano, J.D.~Edelstein, R.A.~Hennigar and
  L.G.~Jaime, \emph{Geometric inflation},
  \href{https://doi.org/10.1016/j.physletb.2020.135242}{\emph{Physics Letters
  B} {\bfseries 802} (2020) 135242}.

\bibitem{Erices:2019mkd}
C.~Erices, E.~Papantonopoulos and E.N.~Saridakis, \emph{{Cosmology in cubic and
  $f(P)$ gravity}},
  \href{https://doi.org/10.1103/PhysRevD.99.123527}{\emph{Phys. Rev. D}
  {\bfseries 99} (2019) 123527}
  [\href{https://arxiv.org/abs/1903.11128}{{\ttfamily 1903.11128}}].

\bibitem{Anson:2019uto}
T.~Anson, E.~Babichev, C.~Charmousis and S.~Ramazanov, \emph{{Cosmological
  instability of scalar-Gauss-Bonnet theories exhibiting scalarization}},
  \href{https://doi.org/10.1088/1475-7516/2019/06/023}{\emph{JCAP} {\bfseries
  06} (2019) 023} [\href{https://arxiv.org/abs/1903.02399}{{\ttfamily
  1903.02399}}].

\bibitem{Antoniou:2020nax}
G.~Antoniou, L.~Bordin and T.P.~Sotiriou, \emph{{Compact object scalarization
  with general relativity as a cosmic attractor}},
  \href{https://doi.org/10.1103/PhysRevD.103.024012}{\emph{Phys. Rev. D}
  {\bfseries 103} (2021) 024012}
  [\href{https://arxiv.org/abs/2004.14985}{{\ttfamily 2004.14985}}].

\bibitem{doi:10.1119/1.17935}
W.F.~Buell and B.A.~Shadwick, \emph{Potentials and bound states},
  \href{https://doi.org/10.1119/1.17935}{\emph{American Journal of Physics}
  {\bfseries 63} (1995) 256}
  [\href{https://arxiv.org/abs/https://doi.org/10.1119/1.17935}{{\ttfamily
  https://doi.org/10.1119/1.17935}}].

\bibitem{Palti:2019pca}
E.~Palti, \emph{{The Swampland: Introduction and Review}},
  \href{https://doi.org/10.1002/prop.201900037}{\emph{Fortsch. Phys.}
  {\bfseries 67} (2019) 1900037}
  [\href{https://arxiv.org/abs/1903.06239}{{\ttfamily 1903.06239}}].

\bibitem{Arkani-Hamed:2015bza}
N.~Arkani-Hamed and J.~Maldacena, \emph{{Cosmological Collider Physics}},
  \href{https://arxiv.org/abs/1503.08043}{{\ttfamily 1503.08043}}.

\bibitem{Adams:2006sv}
A.~Adams, N.~Arkani-Hamed, S.~Dubovsky, A.~Nicolis and R.~Rattazzi,
  \emph{{Causality, analyticity and an IR obstruction to UV completion}},
  \href{https://doi.org/10.1088/1126-6708/2006/10/014}{\emph{JHEP} {\bfseries
  10} (2006) 014} [\href{https://arxiv.org/abs/hep-th/0602178}{{\ttfamily
  hep-th/0602178}}].

\bibitem{inprep}
C.~Erices and S.~Riquelme, \emph{{in preparation}}, .

\bibitem{Meerburg:2016zdz}
P.D.~Meerburg, M.~M\"unchmeyer, J.B.~Mu\~noz and X.~Chen, \emph{{Prospects for
  Cosmological Collider Physics}},
  \href{https://doi.org/10.1088/1475-7516/2017/03/050}{\emph{JCAP} {\bfseries
  03} (2017) 050} [\href{https://arxiv.org/abs/1610.06559}{{\ttfamily
  1610.06559}}].

\end{thebibliography}\endgroup
\end{document}